\documentclass[aps,prd,superscriptaddress,showpacs,preprintnumbers]{revtex4}
\usepackage{amssymb, graphics, setspace}
\usepackage{pgfplots}
\usepackage{textcomp}
\usepackage[caption=false]{subfig}

\usepackage{amsmath}
\usepackage{commath}

\usepackage{graphicx,bm}
\usepackage{url}
\usepackage{float}
\usepackage{textcomp}
\usepackage[labelfont=bf]{caption}
\usepackage{verbatim}
\begin{document}

\newcommand{\dlt}{\bigtriangleup}
\newcommand{\beq}{\begin{equation}}
\newcommand{\eeq}[1]{\label{#1} \end{equation}}
\newcommand{\insertplot}[1]{\centerline{\psfig{figure={#1},width=14.5cm}}}

\parskip=0.3cm


\title{Fine structure of the diffraction cone: manifestation of $t$-channel unitarity}


\author{L\'aszl\'o Jenkovszky}
\affiliation{Bogolyubov Institute for Theoretical Physics (BITP),
Ukrainian National Academy of Sciences \\14-b, Metrologicheskaya str.,
Kiev, 03680, UKRAINE; jenk@bitp.kiev.ua}

\author{Istv\'an Szanyi}
\affiliation{Uzhgorod National University, \\14, Universytets'ka str.,  
Uzhgorod, 88000, UKRAINE; sz.istvan03@gmail.com}

\begin{abstract}
We show that the deviation from exponential behavior of the diffraction cone observed near $t=-0.1$ GeV$^2$ both at the ISR and the LHC (so-called break) follows from a two-pion loop in the $t$-channel, imposed by unitarity. By using a simple Regge-pole model we extrapolate the "break" from the ISR energy region to that of the LHC. 

\end{abstract}


\maketitle

\section{Introduction} \label{s1}

Following TOTEM's impressive results \cite{TOTEM8} on the low-$|t|$ measurements of the $pp$ differential cross section at $8$ TeV, and anticipating their new measurements at $13$ TeV announced recently \cite{TOTEM13} we find it appropriate to remind of the physics behind the observed departure from the exponential behavior of the forward diffraction cone.

For the first time this phenomenon  was observed in 1972 at the the CERN ISR \cite{Bar}, 
a deflection from the exponential behavior of the forward cone in proton-proton scattering 
around $-t=0.1$ GeV$^2$, detected at several energies. 

Experimentalists \cite{TOTEM8, TOTEM13, Bar} quantify the departure from the linear exponential by replacing 
\begin{equation}
|A^N|=a \exp(Bt)\rightarrow a\exp(b_1 t+b_2 t^2+b_3 t^3+...)
\end{equation}
with coefficients $b_i$ fitted to the data. 

This effect can be well fitted \cite{Lia} also by a relevant form factor (residue function) in the Regge-pole scattering amplitude. For a complete and up-to-date review see \cite{Yogi}.

\begin{figure}[ht] 
\centering
\includegraphics[width=.9\textwidth]{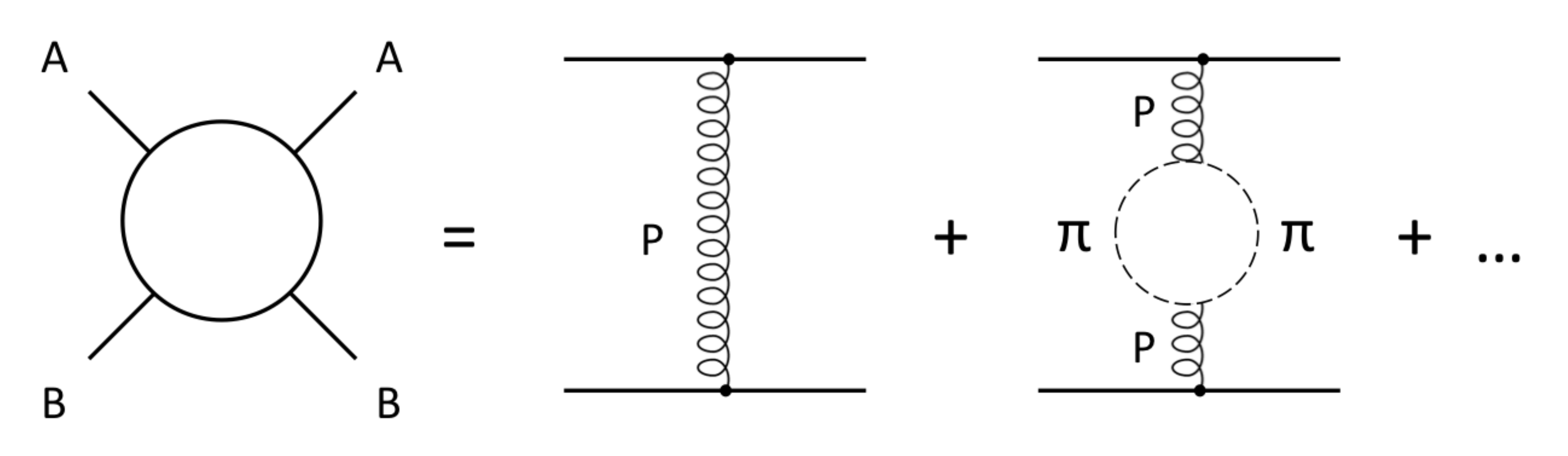}
\caption{Feynman diagram for elastic scattering with a $t$-channel exchange containing a branch point at $t=4m_{\pi}^2$.} 
\label{Fig:Diagram}
\end{figure}

Soon after the ISR measurements, the phenomenon was interpreted \cite{LNC} as manifestation of $t$-channel unitarity, producing a two-pion loop, as
shown in Fig. \ref{Fig:Diagram}, and resulting in a relevant threshold singularity in the Pomeron trajectory. This effect, for bravity called the "break", was confirmed by recent measurements by the TOTEM Collaboration at the CERN LHC, first at $8$ TeV \cite{TOTEM8} and subsequently at $13$ TeV \cite{TOTEM13}.

The new LHC data from TOTEM at $8$ TeV confirm the conclusions made \cite{LNC} about the nature of the break and call for a more detailed analysis of the phenomenon. The new data triggered further theoretical work in this direction \cite{Lengyel, Brazil}, but many issues still remain open. Although the curvature, both at the ISR and the LHC is concave, convex cannot be excluded in other reactions and/or new energies. While the departure from a linear exponential was studied in details both at the ISR and LHC energies, an extra(inter)polation between the two is necessary to clarify the uniqueness of the phenomenon. This is a challenge for the theory, and it can be done within Regge-pole models. Below we do so by using a very simple one, with two Regge exchanges, a Pomeron and a 
secondary effective Reggeon. To test its viability, we first fit its parameters to the proton-proton total section data (Sec. \ref{Sec:Model}).   

The basic premise behind our approach is the introduction of a two-pion loop contribution in the $t$-channel through Regge trajectories, that are non-linear complex functions. As shown by Barut and Zwanziger \cite{Barut}, $t$-channel unitarity constrains the Regge trajectories near the $t$-channel threshold, $t\rightarrow t_0$ by
\begin{equation} \label{Eq:Barut}
\Im \alpha(t)\sim (t-t_0)^{\Re\alpha(t_0)+1/2},
\end{equation} 
where $t_0$ is the lightest threshold, $4m_{\pi}^2$ in the case of the vacuum quantum numbers (Pomeron or $f$ meson). Since $Re\alpha(4m_{\pi}^2)$ is small, a square-root threshold is a reasonable approximation to the above constrain. Higher threshold, inevitable in the trajectory, may be approximated by their power expansion, {\it i.e.} by a linear term, as in 
Eqs. (\ref{Eq:trajectory}). This point is closely related also to the choice of the relevant interval in $t$ under study. Note that the threshold singularity is at positive $t=4m_{\pi}^2$, while the "break" is observed at negative $t$, "symmetric" to $4m_{\pi}^2$. This reflection is a property of analytic functions. The concave departure from the linear exponential, observed in the interval $0 \lesssim |t|\lesssim 0.3$ GeV$^2$ can be fitted by a single square-root threshold in the trajectory, but it would not reproduce the subsequent ($-t\gtrsim 0.3$ GeV$^2$) linearity of the exponential cone, persistent up to the dip (at $\approx -1.4$ GeV$^2$ at the ISR or $-0.6$ GeV$^2$ at the LHC). Note also that we treat only the strong (nuclear) amplitude, separated from Coulombic forces.
Thus, the "break" (in fact a smooth deflection of the linear exponential) of the cone, has a relatively narrow location around $-t \approx 0.1 \pm 0.01$ GeV$^2,$ both at the ISR and the LHC energies, whereupon it recovers its exponential shape, followed by the dip, whose position is strongly energy-dependent. 

In the present paper we study the "break" within a simple Regge-pole model, assuming the universality of this phenomenon in high-energy hadron scattering. On fitting the model to the data, we proceed in two ways: 1) trying to minimize the number of free parameters, we adopt standard values for the trajectories, {\it e.g.} from Ref. \cite{Landshoff}; 2) on the other hand, in view of the oversimplified nature of our "effective" Regge-pole model, we optionally let these parameters free, fitting them to the data.       

In Sec. \ref{Sec:Model} we introduce a simple Regge-pole model, normalizing its energy dependence to $pp$ total cross section data. In \ref{Sec:ISR} we revise the 1972 ISR data and their fits to a Regge-pole model with a $t$-channel threshold imposed by unitarity. In Sec. \ref{Sec:LHC} a similar analyses of the LHC data is presented. Central is Sec. \ref{Extrapolate}, in which, by Regge-extrapolating the cross section from the ISR energy region to that of the ISR, we map the "break" fitted at the ISR to that seen at the LHC. Some conclusions are drawn in Sec. \ref{Sec:Conclude}.      

\section{A simple Regge-pole model}\label{Sec:Model}
For our purposes we use a simple Regge pole model with a supercritical Pomeron \cite{Landshoff} and an effective Reggeon contributions, denoted by $A_f$, close (but not similar) to the $f$ Reggeon, 
\begin{equation} \label{Eq:ampl}
A(s,t)=A_P(s,t)+A_f(s,t),
\end{equation} 
where
\begin{equation}\label{Eq:Pf}
A_P(s,t)=-a_Pe^{b_P\alpha_P(t)}e^{-i\pi\alpha_P(t)/2}(s/s_{0P})^{\alpha_P(t)}, \ \ \ A_f(s,t)=-a_fe^{b_f\alpha_f(t)}e^{-i\pi\alpha_f(t)/2}(s/s_{0f})^{\alpha_f(t)},
\end{equation}
with the trajectories
\begin{equation}\label{Eq:trajectory}
\alpha_P(t)=\alpha_{0P}+\alpha'_Pt-\alpha_{1P}(\sqrt{4m_{\pi}^2-t}-2m_{\pi}), \ \ \ \alpha_f(t)=\alpha_{0f}+\alpha'_ft-\alpha_{1f}(\sqrt{4m_{\pi}^2-t}-2m_{\pi}).
\end{equation}
We use the norm:
\begin{equation}
\sigma_T(s)=\frac{4\pi}{s}\Im A(s,t=0),\  \  \ \frac{d\sigma}{dt}=\frac{\pi}{s^2}|A(s,t)|^2.
\end{equation}
The model contains 12 free parameters ($a_P$ ($\sqrt{mbGeV^2}$), $b_P$ (dimensionless), $\alpha_{0P}$ (dimensionless), $\alpha'_P$ (GeV$^{-2}$), $\alpha_{1P}$ (GeV$^{-1}$), $s_{0P}$ (GeV$^2$), $a_f$ ($\sqrt{mbGeV^2}$), $b_f$ (dimensionless), $\alpha_{0f}$ (dimensionless), $\alpha'_f$ (GeV$^{-2}$), $\alpha_{1f}$ (GeV$^{-1}$),  $s_{0f}$ (GeV$^2$)), most of which are known a priori, needing only fine-tuning. We shall optionally use known values of the parameters and/or let them free.  

Anticipating detailed fits to the low-$|t|$ data, we start with a simple fit to the data on proton-proton total cross section starting from  2.3 GeV, with fixed intercepts of the Pomeron 
$\alpha(0)_P=1.0808$ and of the effective Reggeon $\alpha(0)_f=0.5$ \cite{Landshoff}. From this fit we find the parameters $a_P$ and $a_f$, to be fine-tuned in what follows. We found by trial that the scaling parameters $s_0$ do not affect significantly the resulting fits, so we set $s_0=1$ (GeV$^2$) everywhere, both for the Pomeron and for the $"f"$. 

\begin{figure}[H]
 	\centering
 	\includegraphics[scale=0.24]{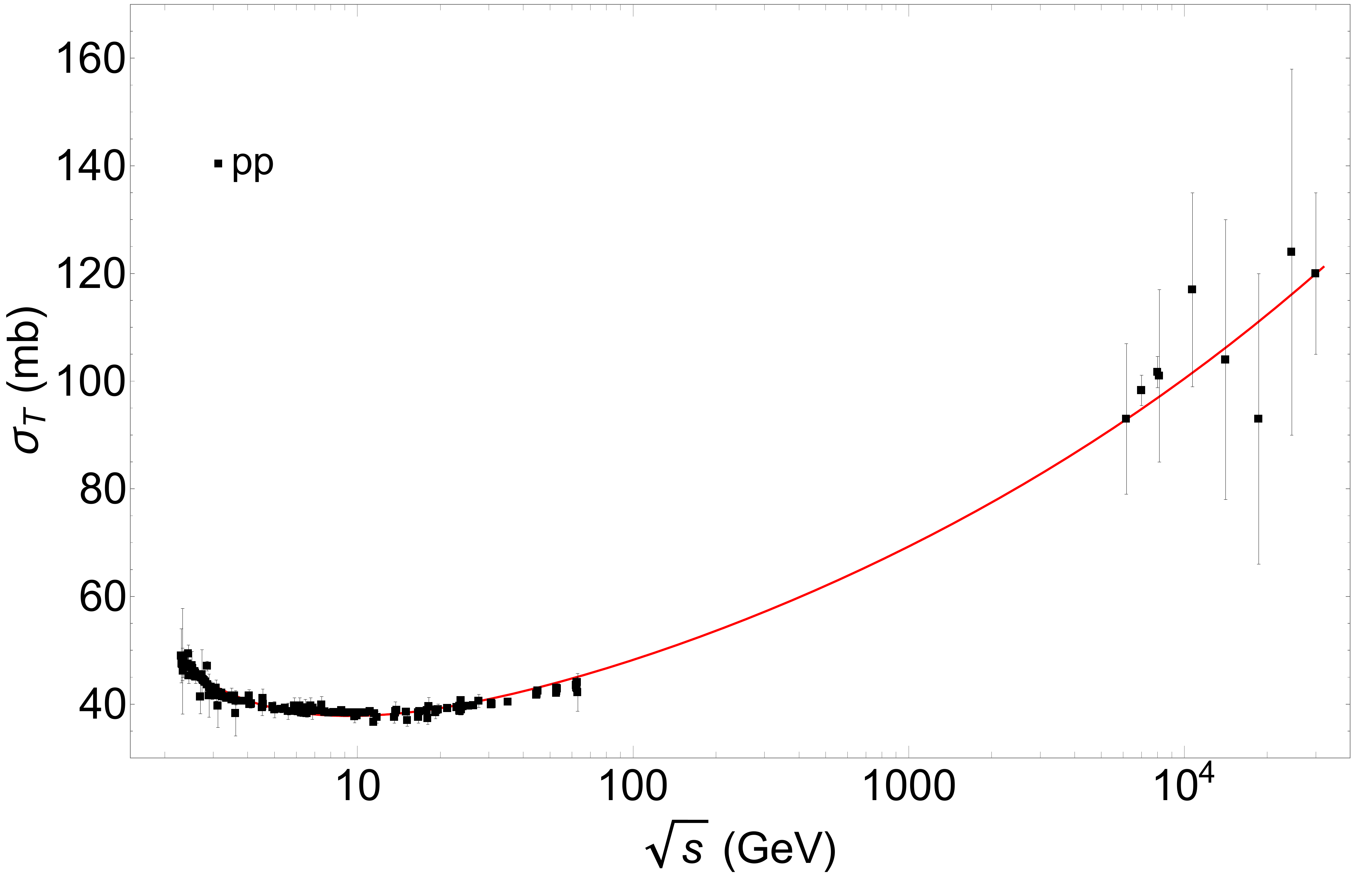}
 	\caption{Fit to $pp$ total cross section with fixed intercepts. The data are from \cite{ISR}}
 	\label{Fig:Total1}
 \end{figure}

The resulting fit is shown in Fig. \ref{Fig:Total1}.
 
The fit gives $a_P=1.44411$, $a_f=1.56448$, $b_P=1.08699$, $b_f=4.42013$, thus the total cross section is determined by the expression: $\sigma_T(s)=22.6709s^{0.0808}+49.2985s^{-0.5}$.

\begin{figure}[H]
	\centering
	\includegraphics[scale=0.24]{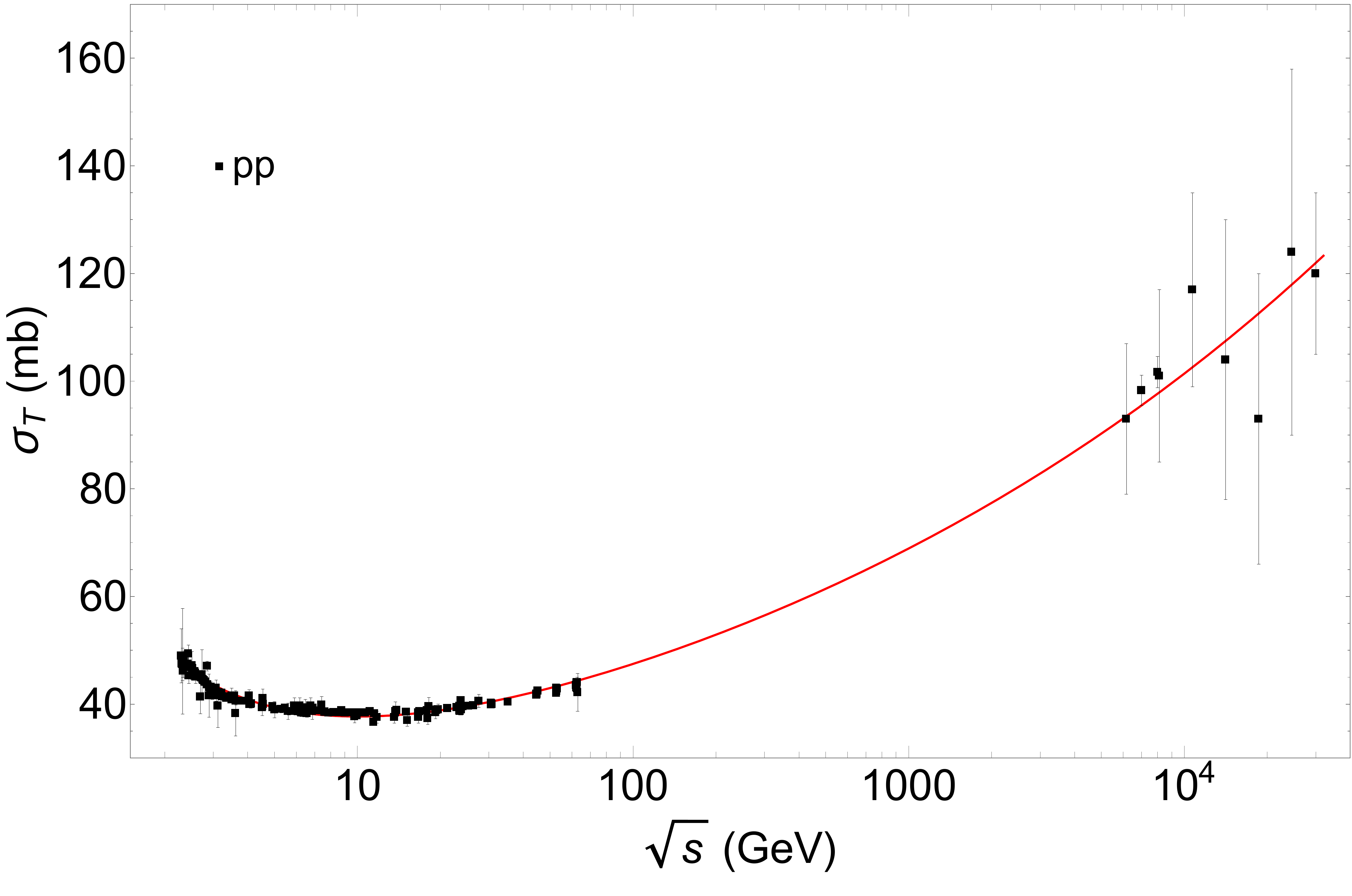}
	\caption{Fit to $pp$ total cross section without fixed intercepts. The data are from \cite{ISR}}
	\label{Fig:Total2}
\end{figure}

If we do not fix the values of $\alpha_{0P}$ and $\alpha_{0f},$ we obtain a slightly different fit for the total cross section of pp scattering shown in Fig. \ref{Fig:Total2}. The values of fitted parameters:
$\alpha_{0P}$=1.08414, $\alpha_{0f}$=0.550223, $a_P$=1.21022, $a_f$=2.20749, $b_P$=1.1991, $b_f$=3.19442. In this case the total cross section is determined as $\sigma_T(s)=21.5179s^{0.0841}+47.5958s^{-0.4498}$.

\section{The "break" at ISR} \label{Sec:ISR}  
At the ISR the proton-proton differential cross section was measured at $\sqrt {s}=23.5, 30.7, 44.7, 52.8$ and $62.5$ GeV in the interval $0.01<-t<0.35$ GeV$^2$. In all the above energy intervals the differential cross section changes its slope near $-t=0.1$ GeV$^2$ by about two units of GeV$^2$. Below we fit the ISR data to a simple Regge pole model with two Regge exchanges - the Pomeron and an effective sub-leading trajectory.

\begin{figure}[H] 
	\centering
	\includegraphics[scale=0.19]{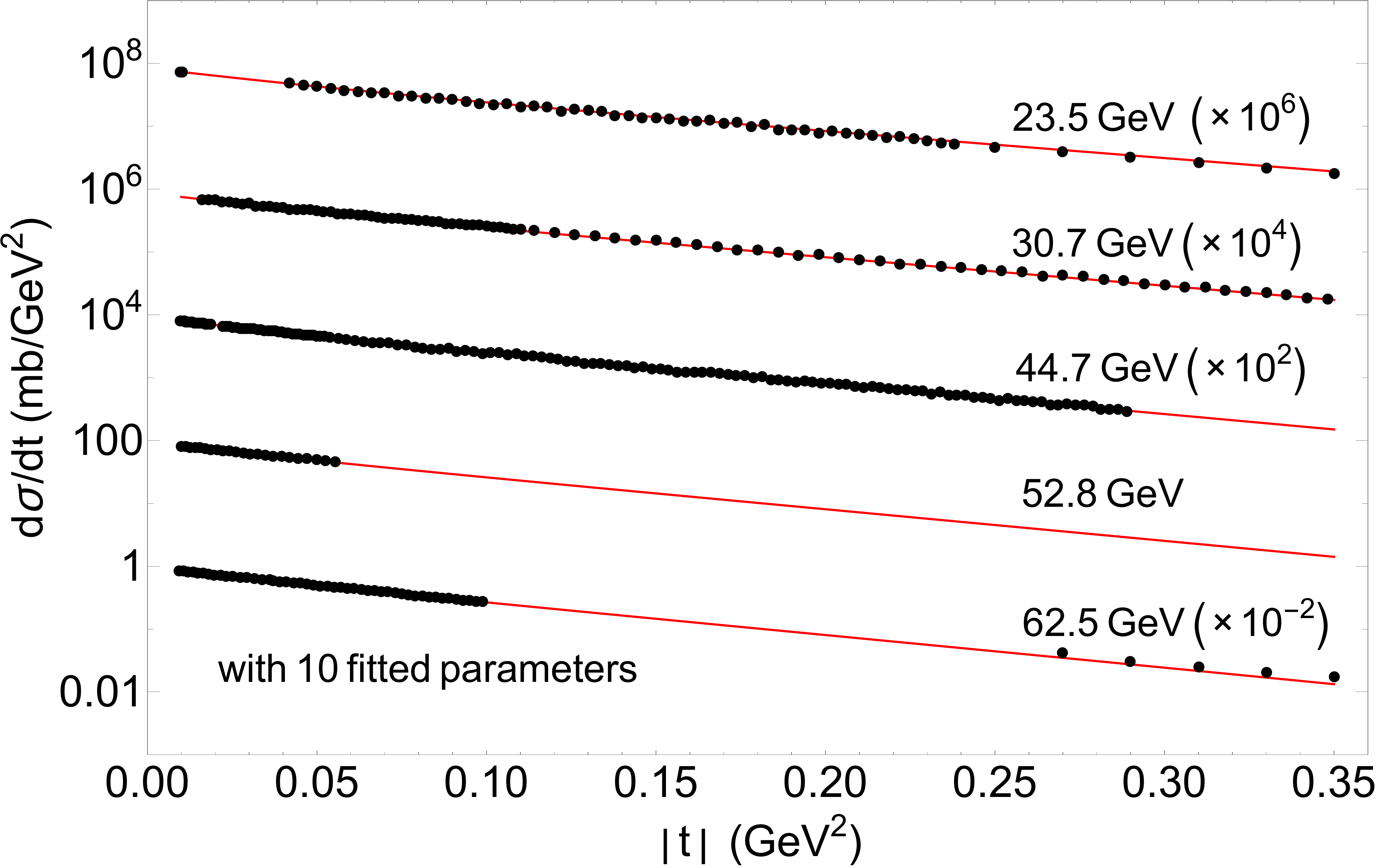}
	\qquad
	\includegraphics[scale=0.19]{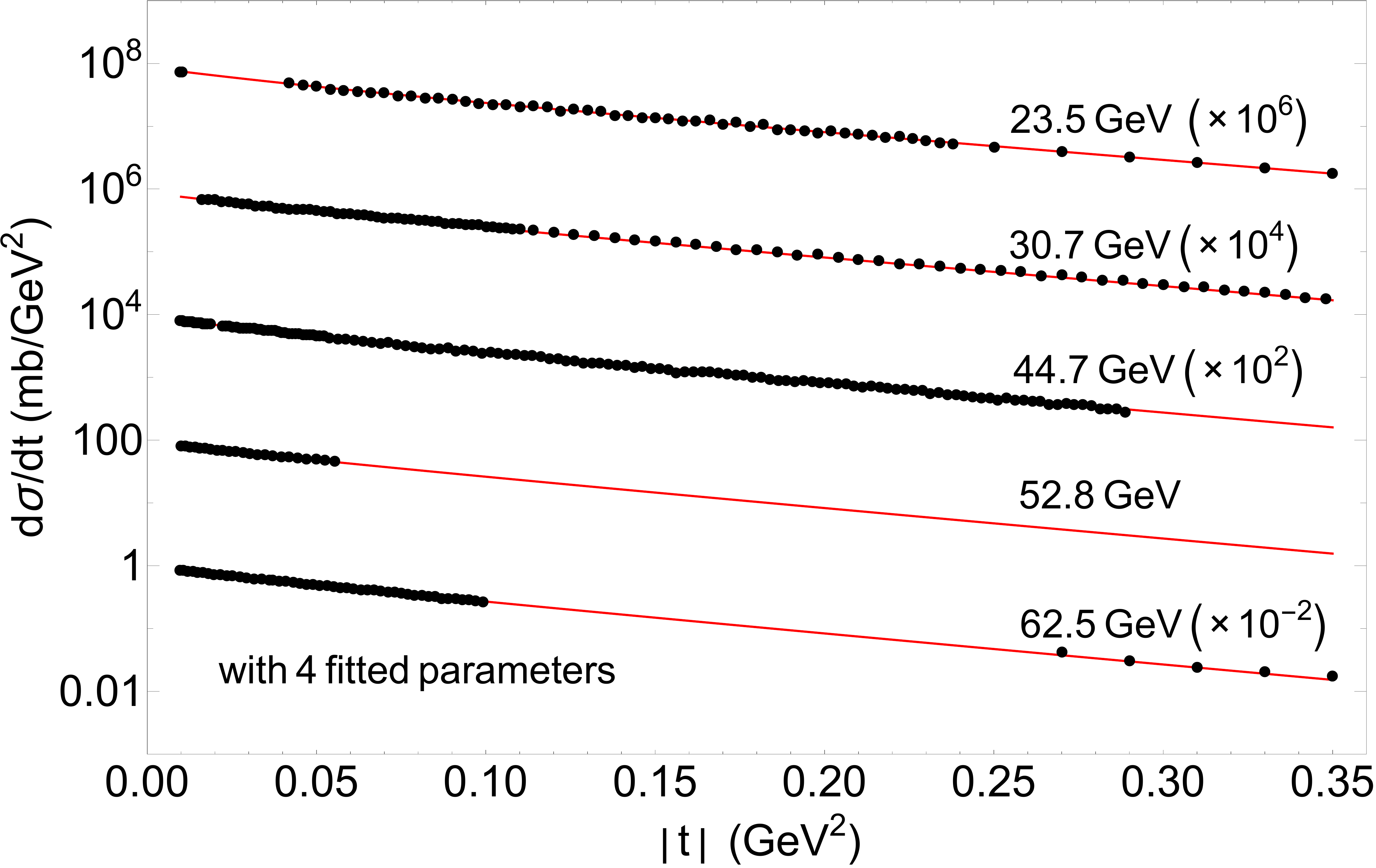}
	\caption{Present fit to the ISR data \cite{ISR}.}
	\label{Fig:ISR}
\end{figure}

The result of the fit is shown in Fig. \ref{Fig:ISR}.

The values of the fitted parameters are presented in Table I.
\begin{table}[H]
	\centering
	\subfloat[With 10 fitted parameters \label{sfig:testa}]{%
		\begin{tabular}{c c c c}\hline
			$\alpha_{0P}$& 1.11828&	$\alpha_{0f}$& 0.817835\\
			$\alpha'_P$& 0.788847&$\alpha'_f$& 0.814786\\
			$\alpha_{1P}$& -0.189332&	$\alpha_{1f}$&0.290017\\
			$a_P$&0.819335&$a_f$&0.158121\\
			$b_P$&1.04319&$b_f$&4.99955\\
			$s_{0P}$&1 (fixed)&$s_{0f}$&1 (fixed)\\\hline
			&$\chi^2/DOF$ & 0.5251&\\
			&$DOF$ & 353& \\\hline
		\end{tabular}%
	}\qquad
	\subfloat[With 4 fitted parameters \label{sfig:testa}]{%
		\begin{tabular}{c c c c}\hline
			$\alpha_{0P}$& 1.08 (fixed)&	$\alpha_{0f}$& 0.5 (fixed)\\
			$\alpha'_P$& 0.3 (fixed)&$\alpha'_f$& 1 (fixed)\\
			$\alpha_{1P}$& 0.03 (fixed)&	$\alpha_{1f}$&0.1 (fixed)\\
			$a_P$&0.000223008&$a_f$&0.140832\\
			$b_P$&9.148649&$b_f$&11.3814\\
			$s_{0P}$&1 (fixed)&$s_{0f}$&1 (fixed)\\\hline
			&$\chi^2/DOF$ &0.5746&\\
			&$DOF$ & 359& \\\hline
		\end{tabular}%
	}
	
	\caption{Values of the fitted parameters for ISR energies \cite{ISR}.}
	\label{Table.1}
\end{table}

\begin{figure}[H] 
	\centering
	\subfloat[23.5 GeV\label{sfig:testa}]{%
		\includegraphics[scale=0.19]{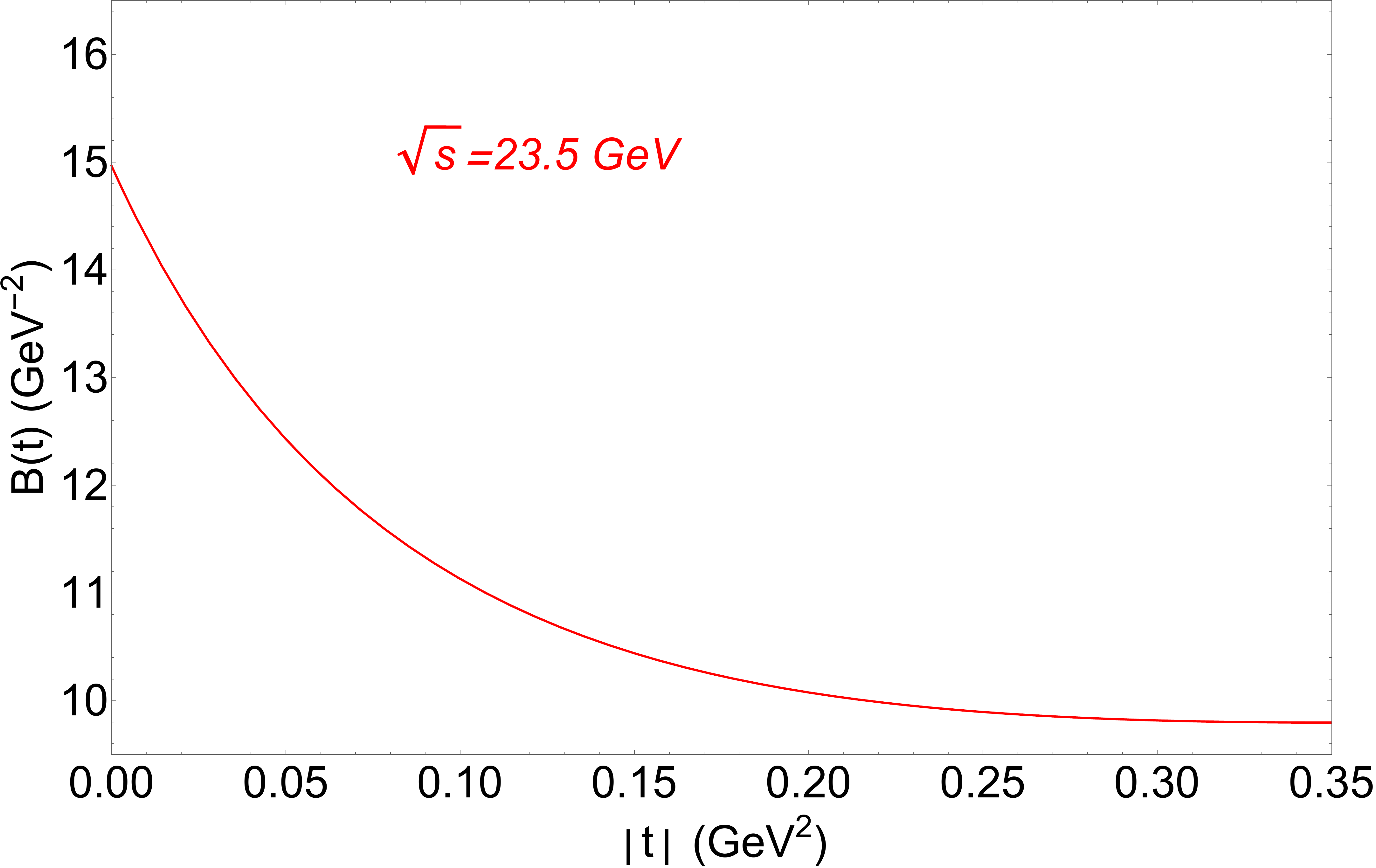}%
	}\hfill
	\subfloat[30.7 GeV\label{sfig:testa}]{%
		\includegraphics[scale=0.19]{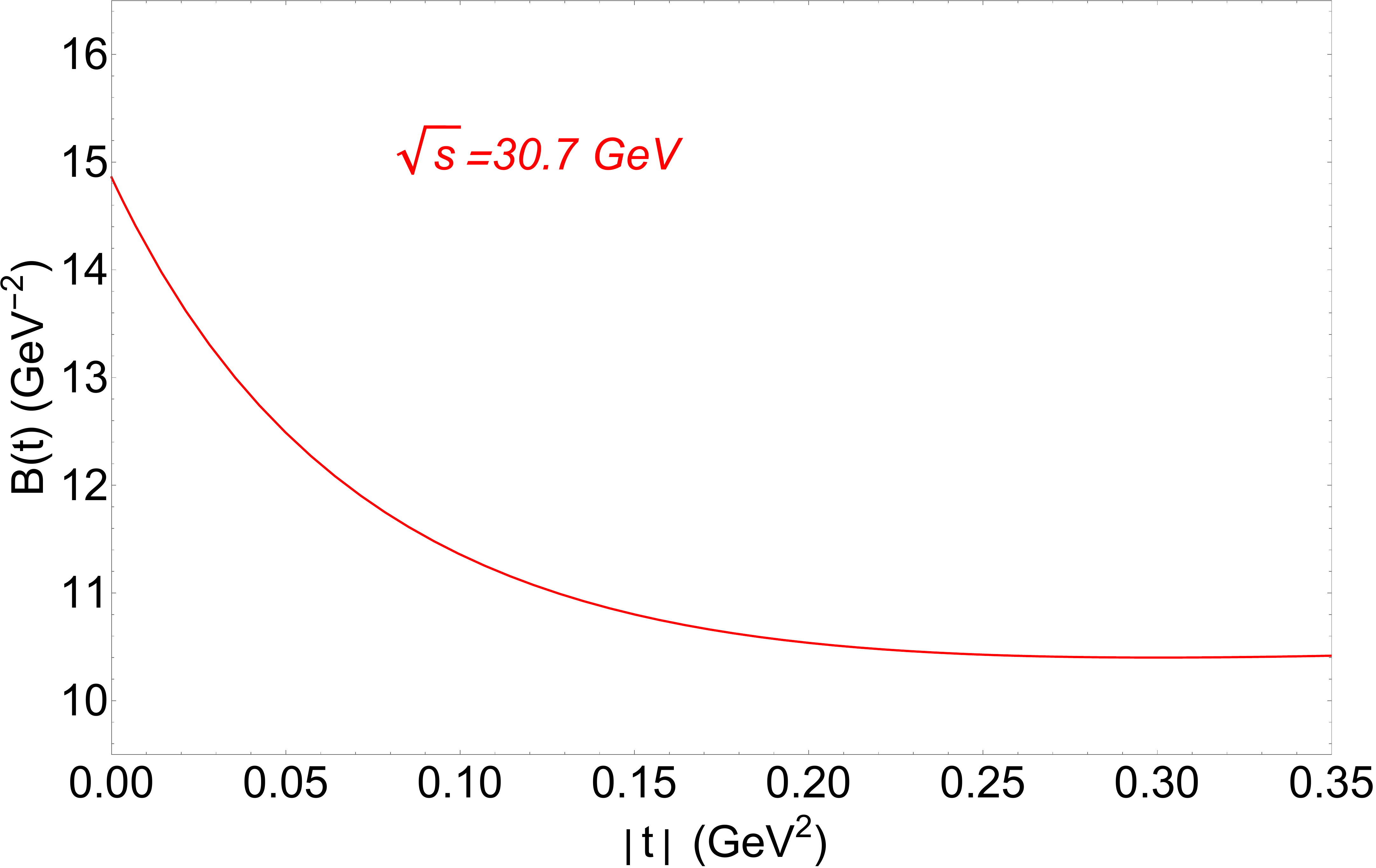}%
	}\hfill
	\subfloat[44.7 GeV\label{sfig:testa}]{%
		\includegraphics[scale=0.19]{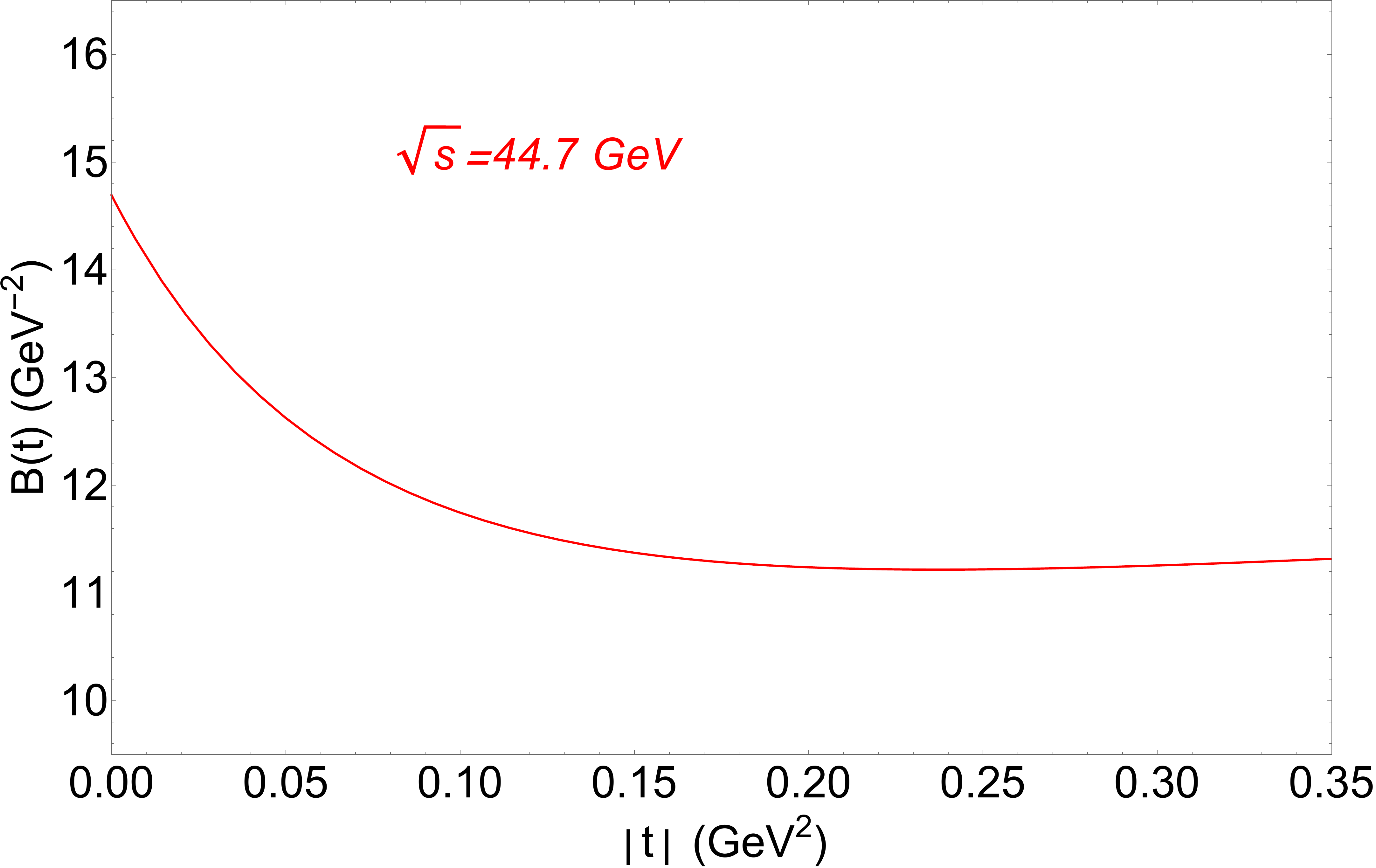}%
	}\hfill
	\subfloat[52.8 GeV\label{sfig:testa}]{%
		\includegraphics[scale=0.19]{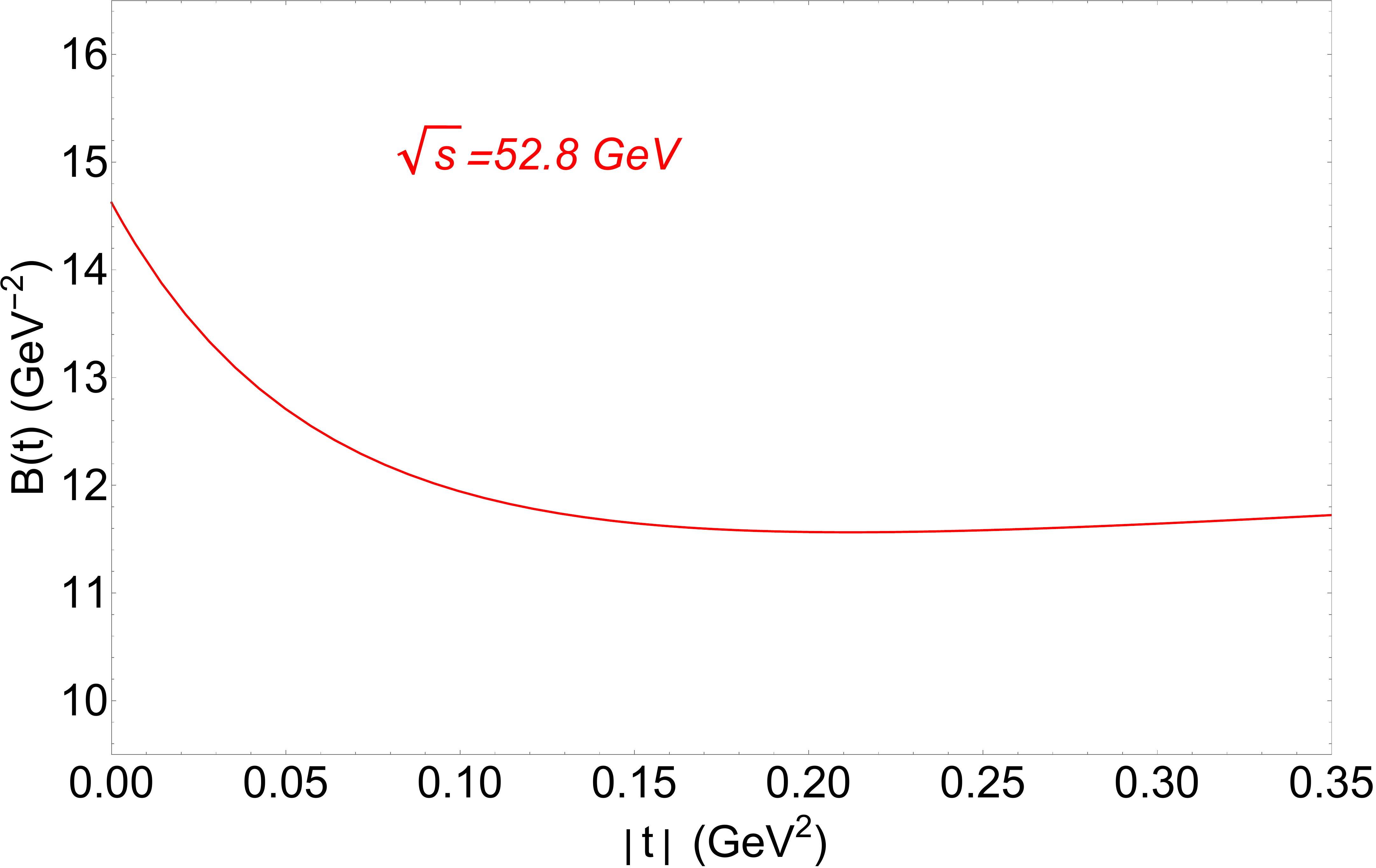}%
	}\hfill
	\subfloat[62.5 GeV\label{sfig:testa}]{%
		\includegraphics[scale=0.19]{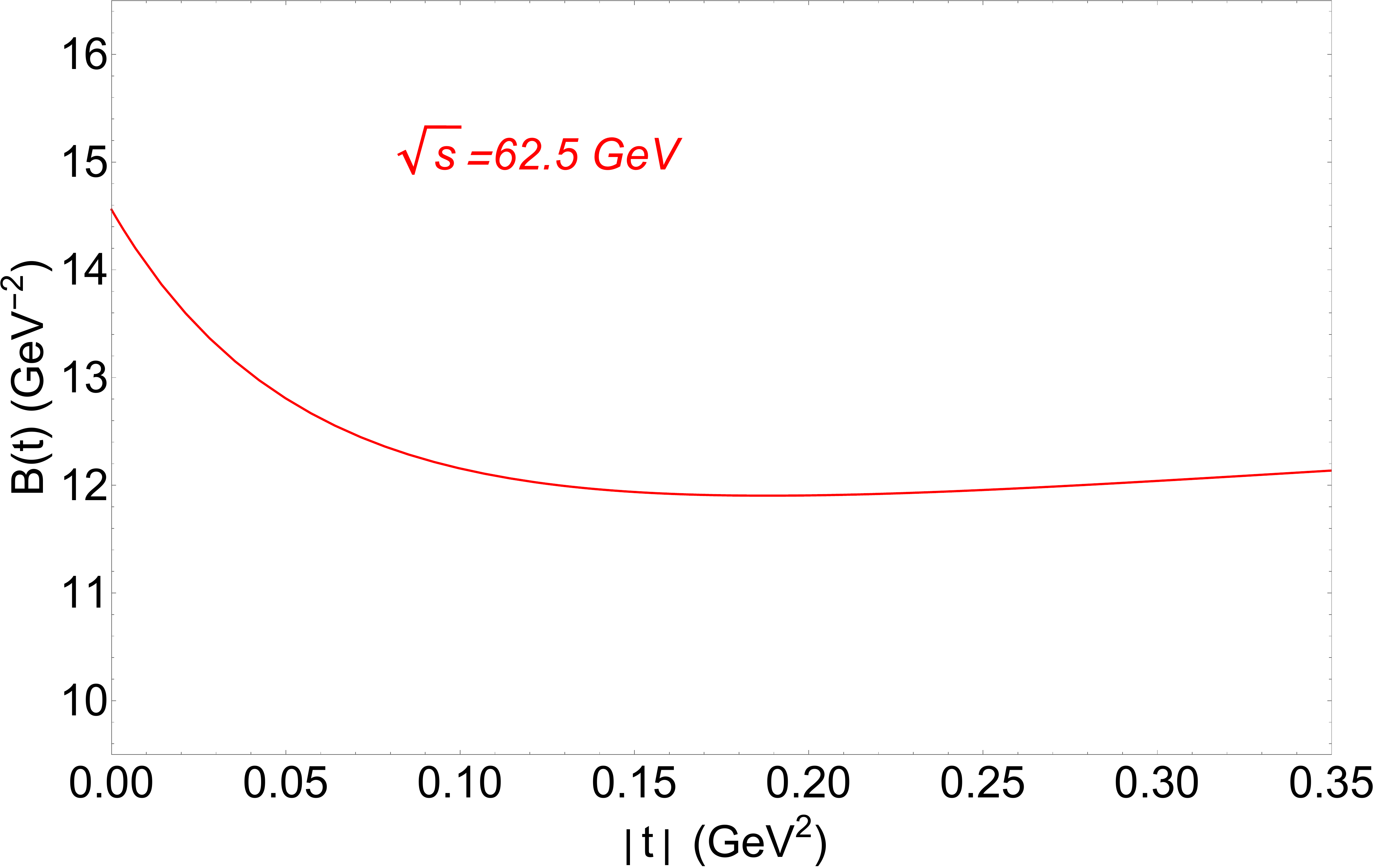}%
	}
	
	\caption{Local slopes calculated for ISR energies \cite{ISR} with $10$ fitted parameters.}
	\label{Fig:ISR_slopea}
\end{figure}
\begin{figure}[H] 
	\centering
	\subfloat[23.5 GeV\label{sfig:testb}]{%
		\includegraphics[scale=0.19]{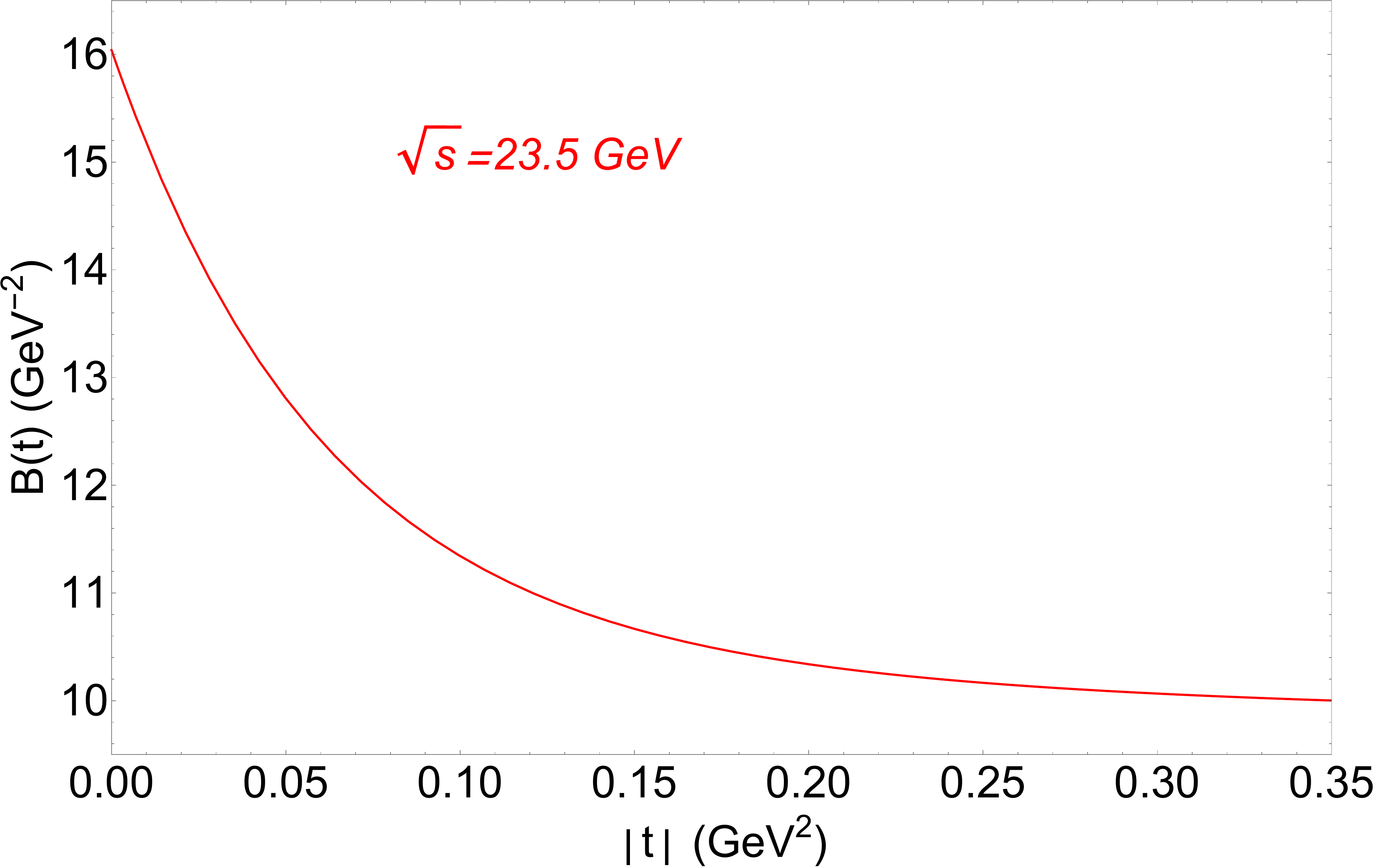}%
	}\hfill
	\subfloat[30.7 GeV\label{sfig:testb}]{%
		\includegraphics[scale=0.19]{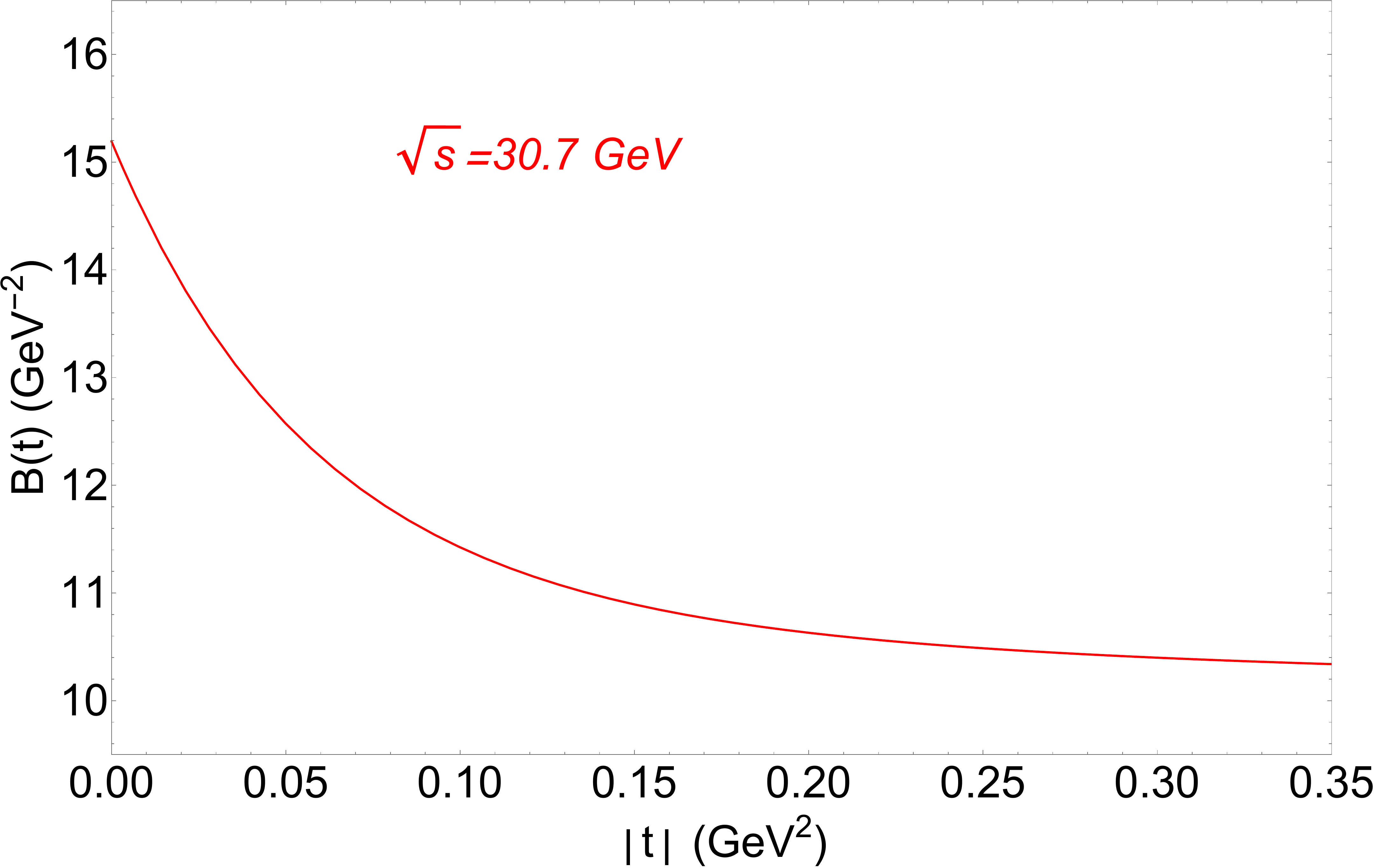}%
	}\hfill
	\subfloat[44.7 GeV\label{sfig:testb}]{%
		\includegraphics[scale=0.19]{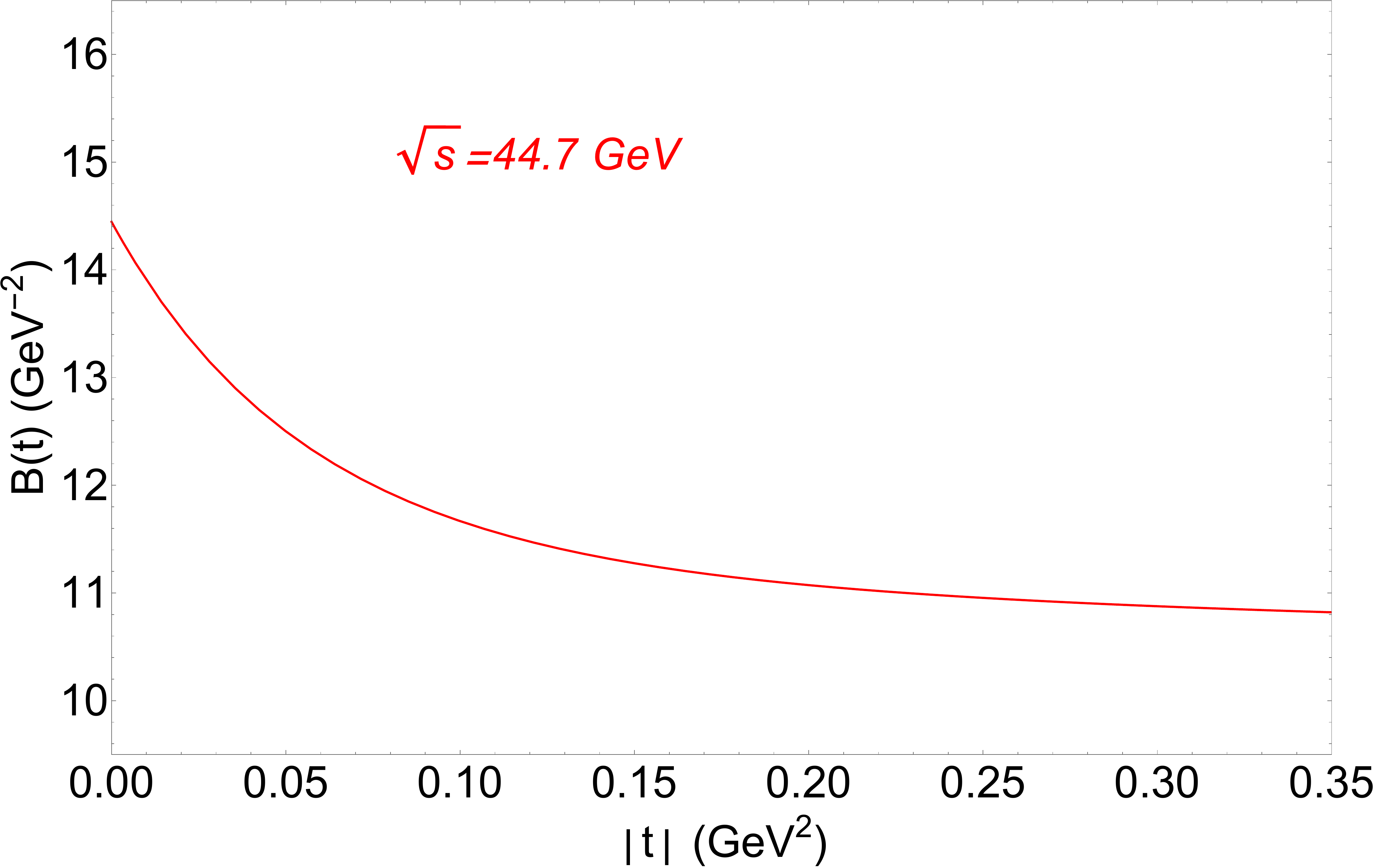}%
	}\hfill
	\subfloat[52.8 GeV\label{sfig:testb}]{%
		\includegraphics[scale=0.19]{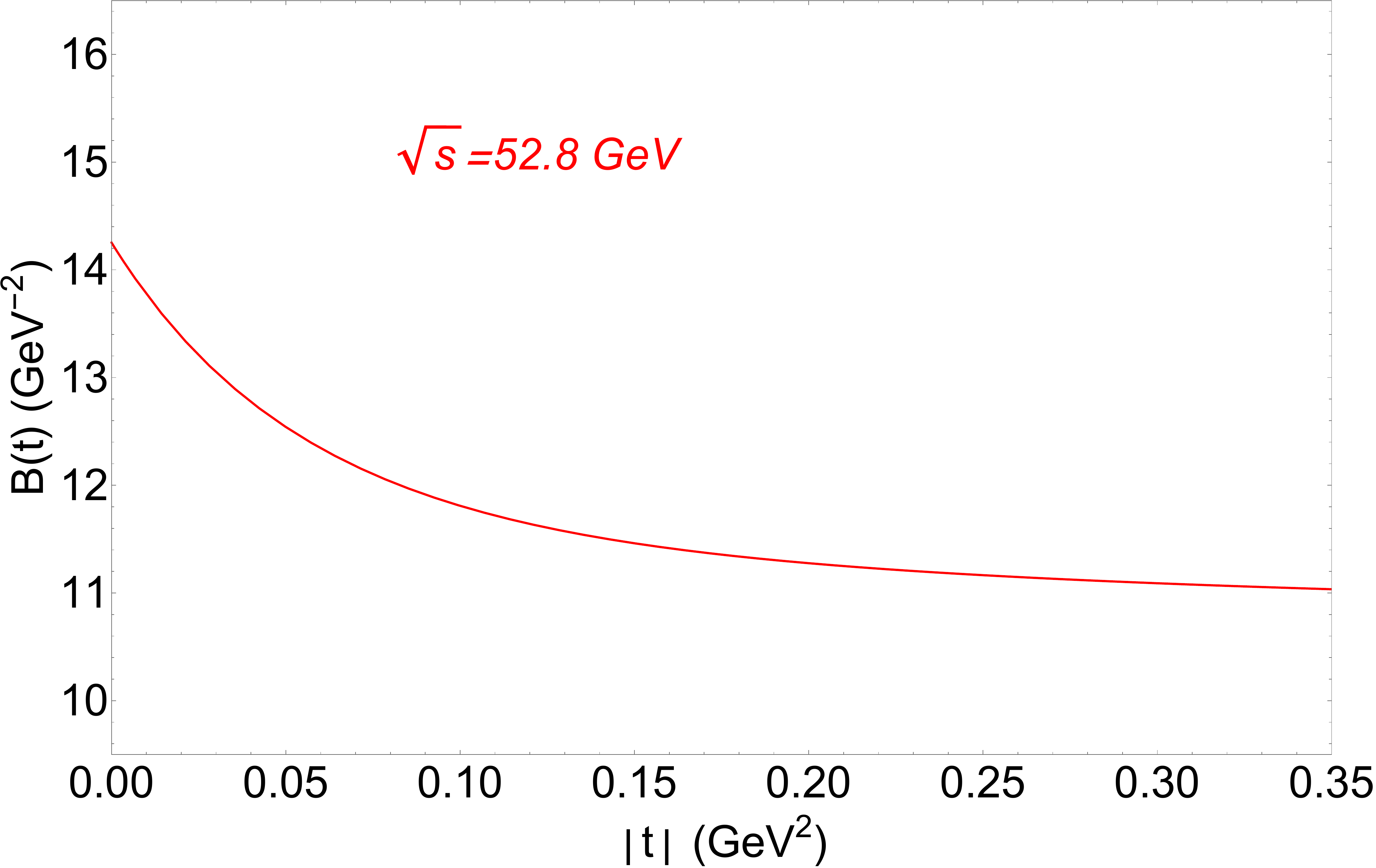}%
	}\hfill
	\subfloat[62.5 GeV\label{sfig:testb}]{%
		\includegraphics[scale=0.19]{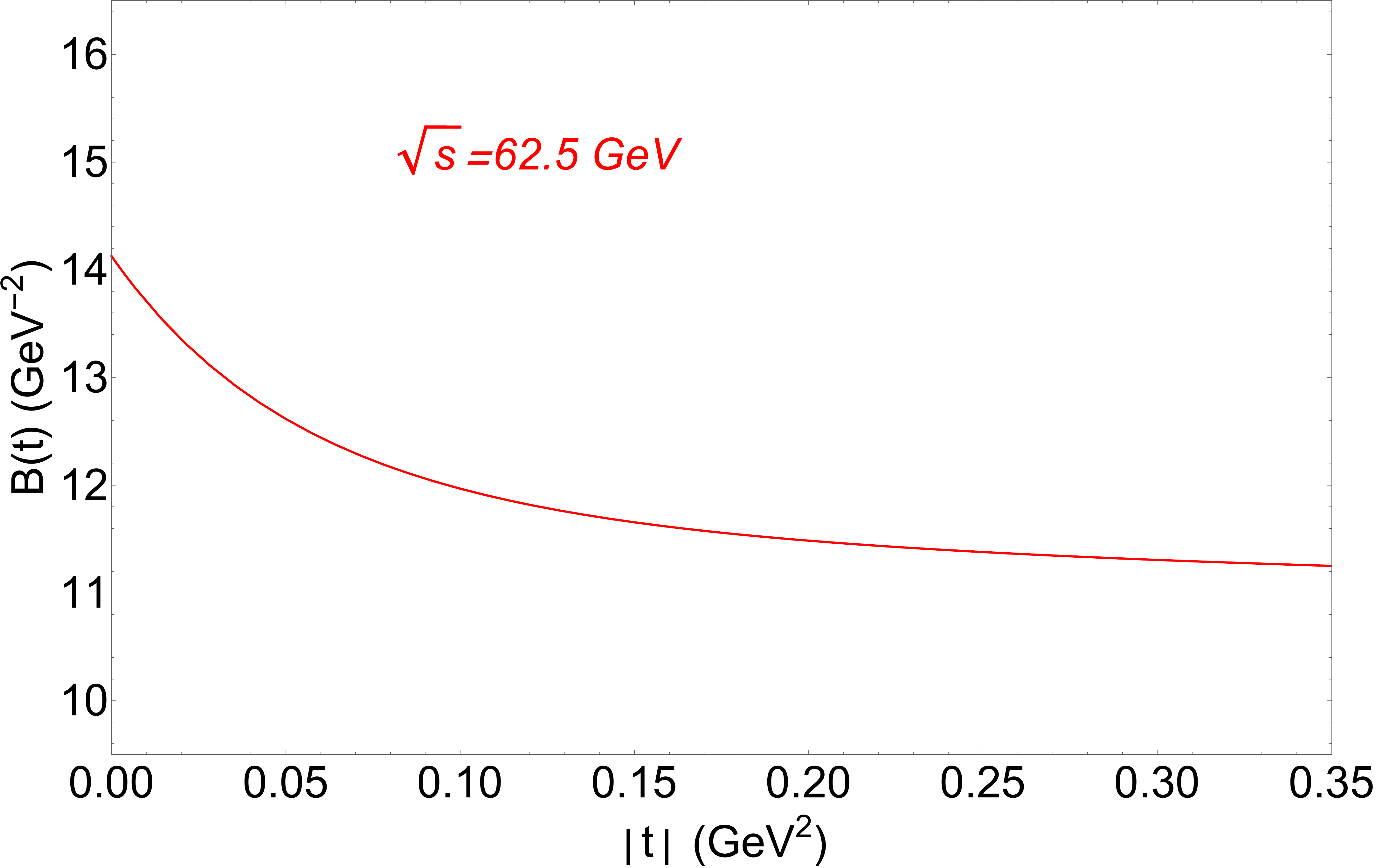}%
	}
	
	\caption{Local slopes calculated for ISR energies \cite{ISR} with $4$ fitted parameters.}
	\label{Fig:ISR_slopeb}
\end{figure}
\begin{figure} 
	\centering
	\subfloat[23.5 GeV
	\label{sfig:testa}]{%
		\includegraphics[scale=0.19]{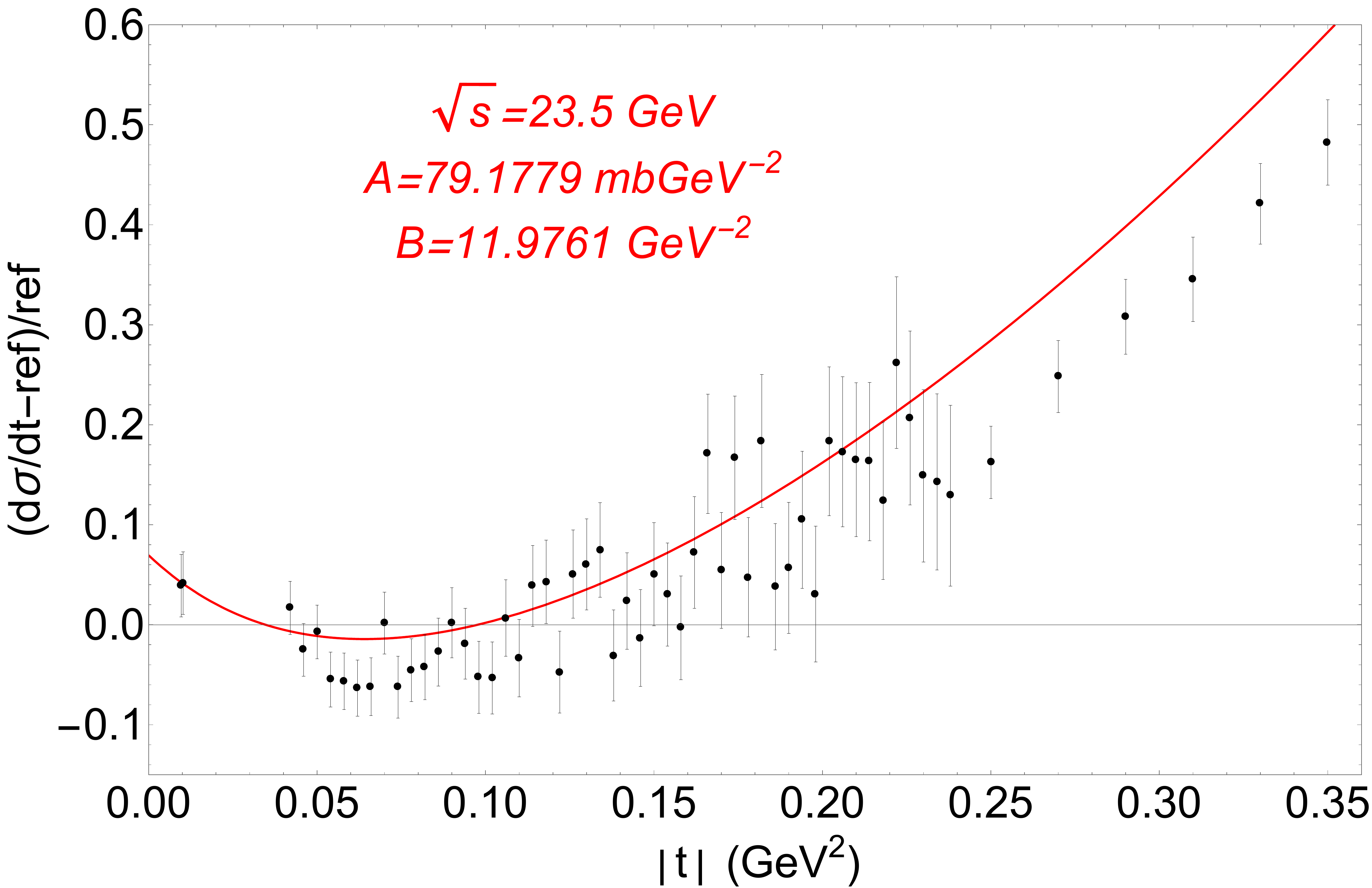}%
	}\hfill
	\subfloat[30.7 GeV\label{sfig:testa}]{%
		\includegraphics[scale=0.19]{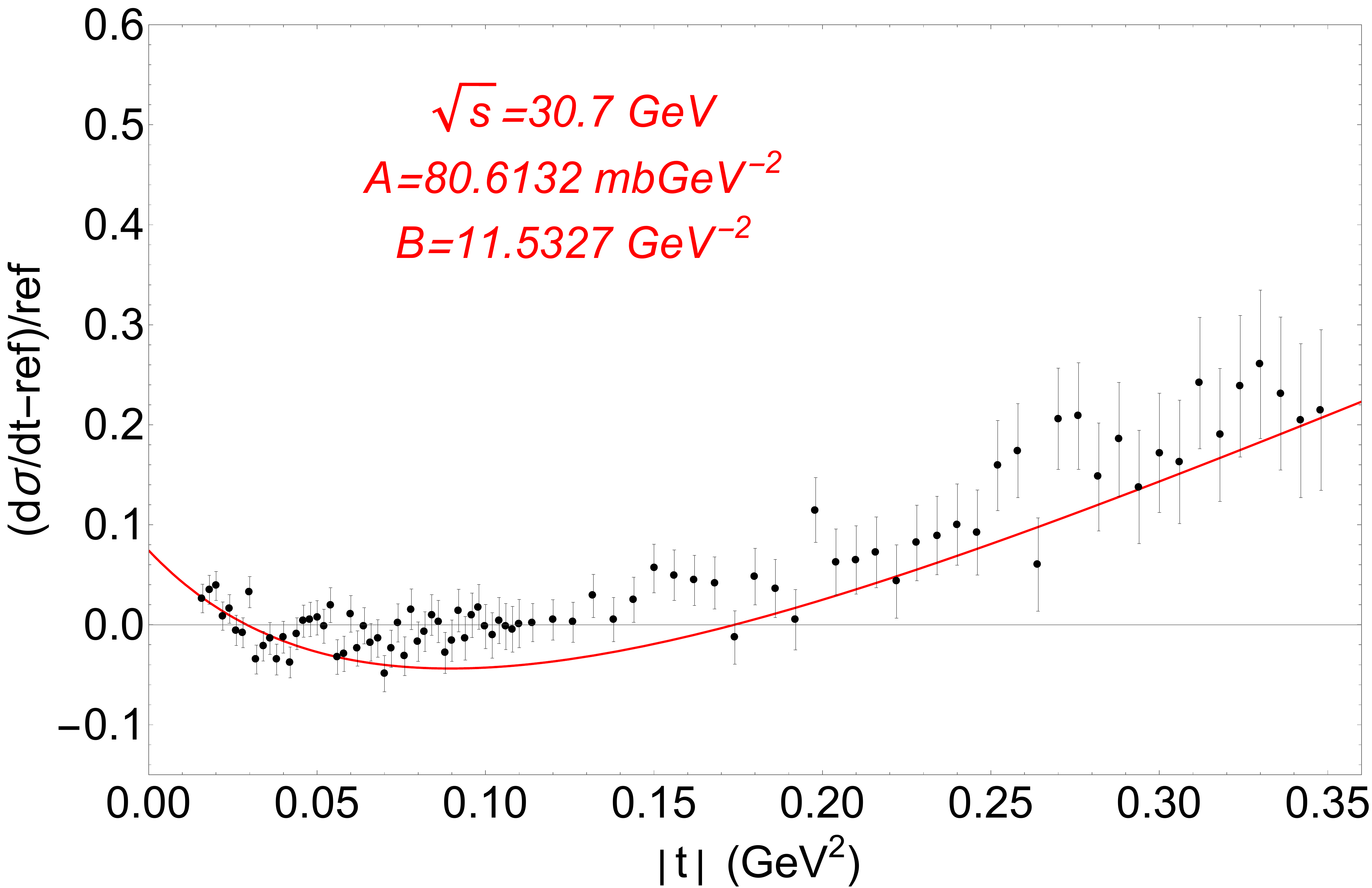}%
	}\hfill
	\subfloat[44.7 GeV\label{sfig:testa}]{%
		\includegraphics[scale=0.19]{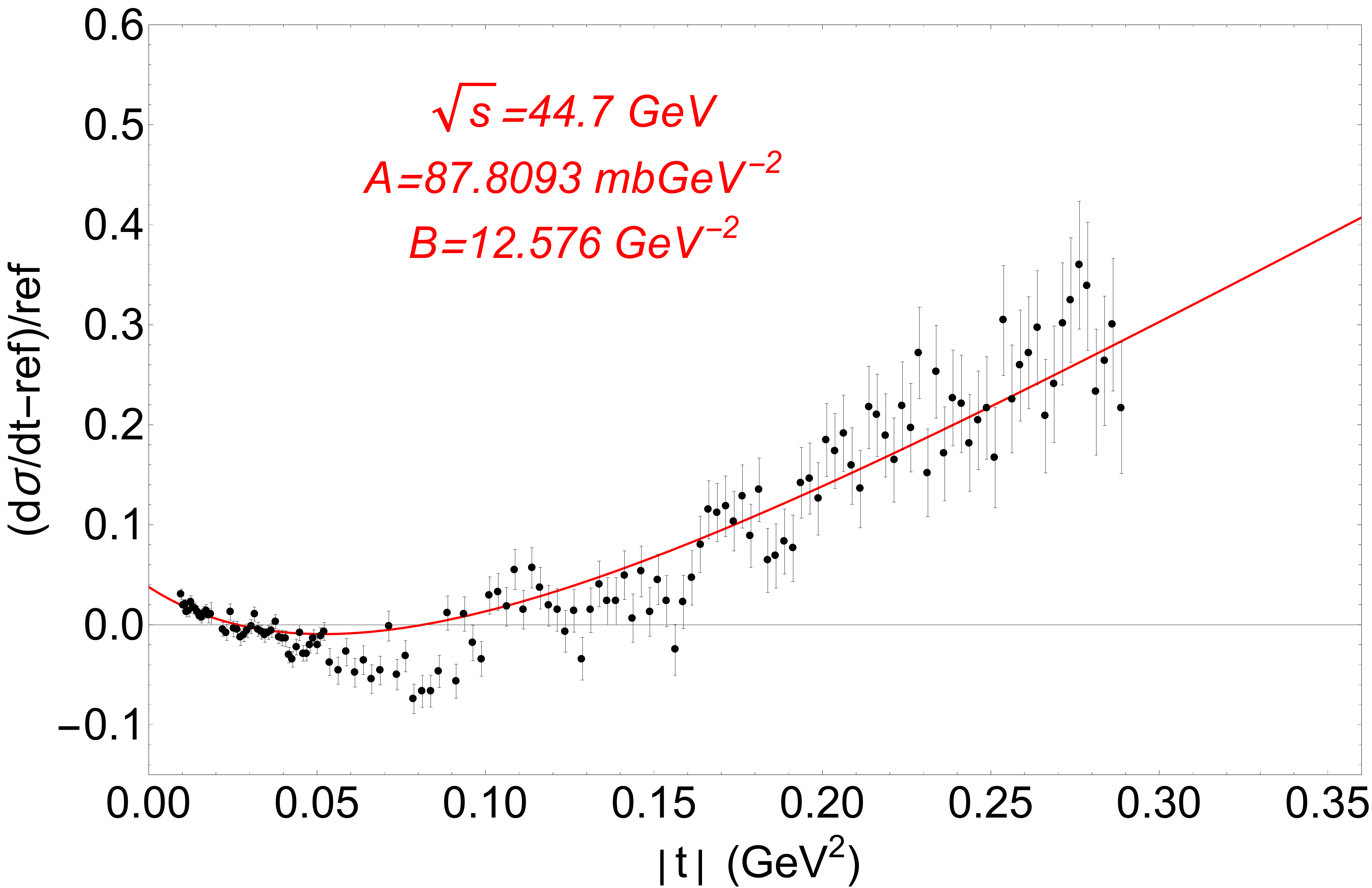}%
	}\hfill
	\subfloat[52.8 GeV\label{sfig:testa}]{%
		\includegraphics[scale=0.19]{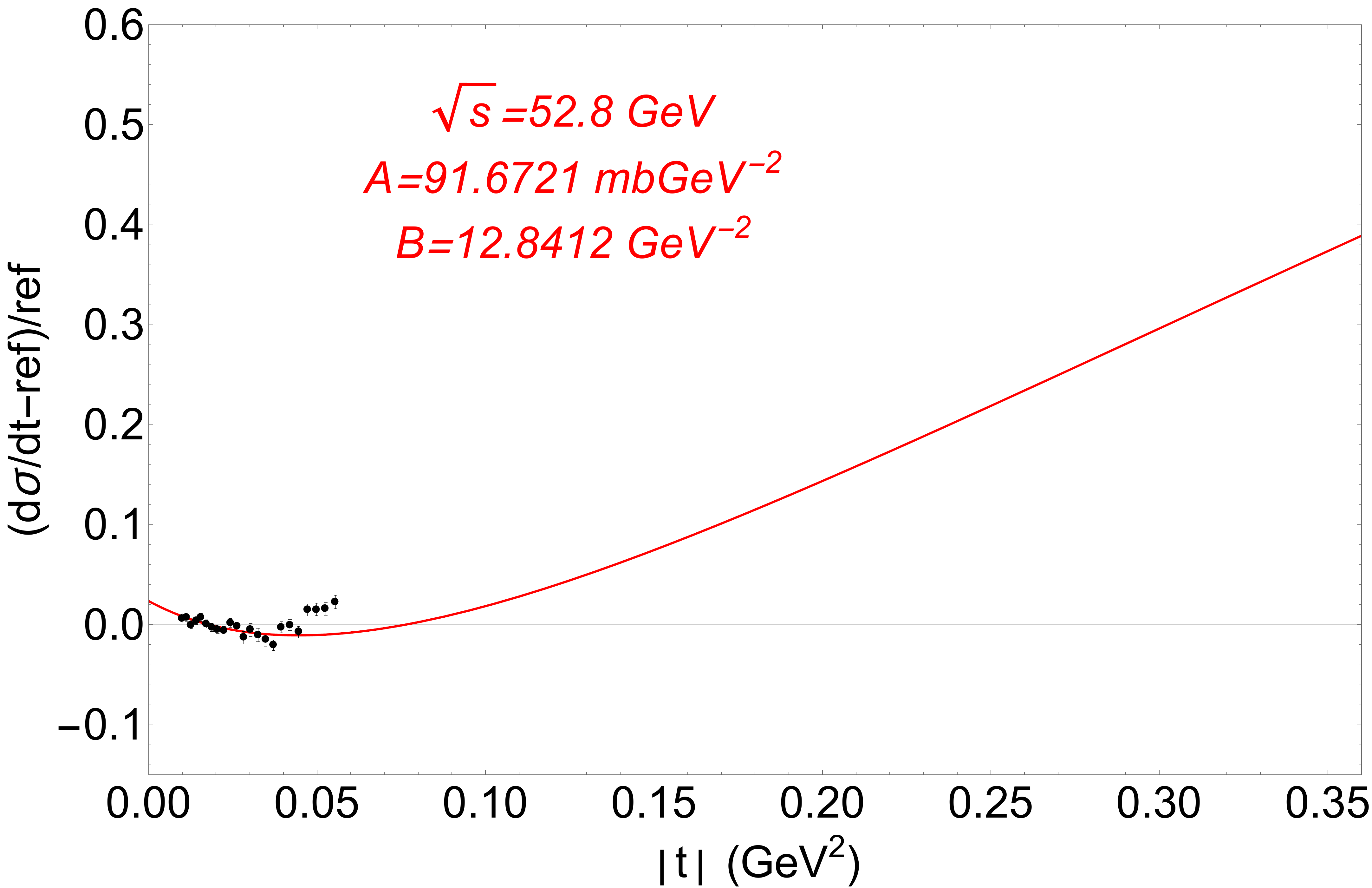}%
	}\hfill
	\subfloat[62.5 GeV\label{sfig:testa}]{%
		\includegraphics[scale=0.19]{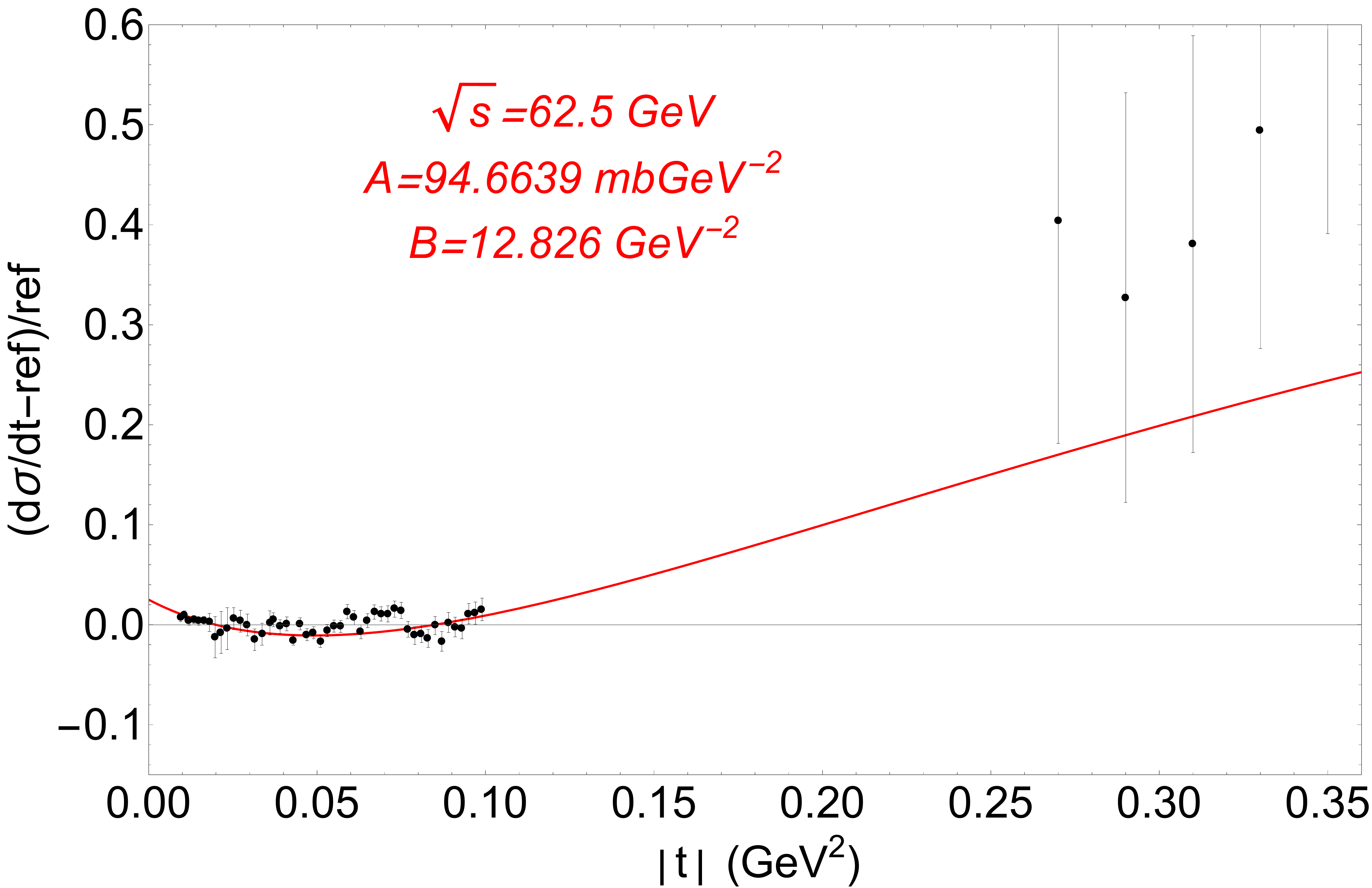}%
	}
	
	\caption{R ratios calculated for the ISR \cite{ISR} energies with $10$ fitted parameters.}
	\label{Fig:ISR_norma}
\end{figure}
\begin{figure}[H]  
	\centering
	\subfloat[23.5 GeV
	\label{sfig:testa}]{%
		\includegraphics[scale=0.19]{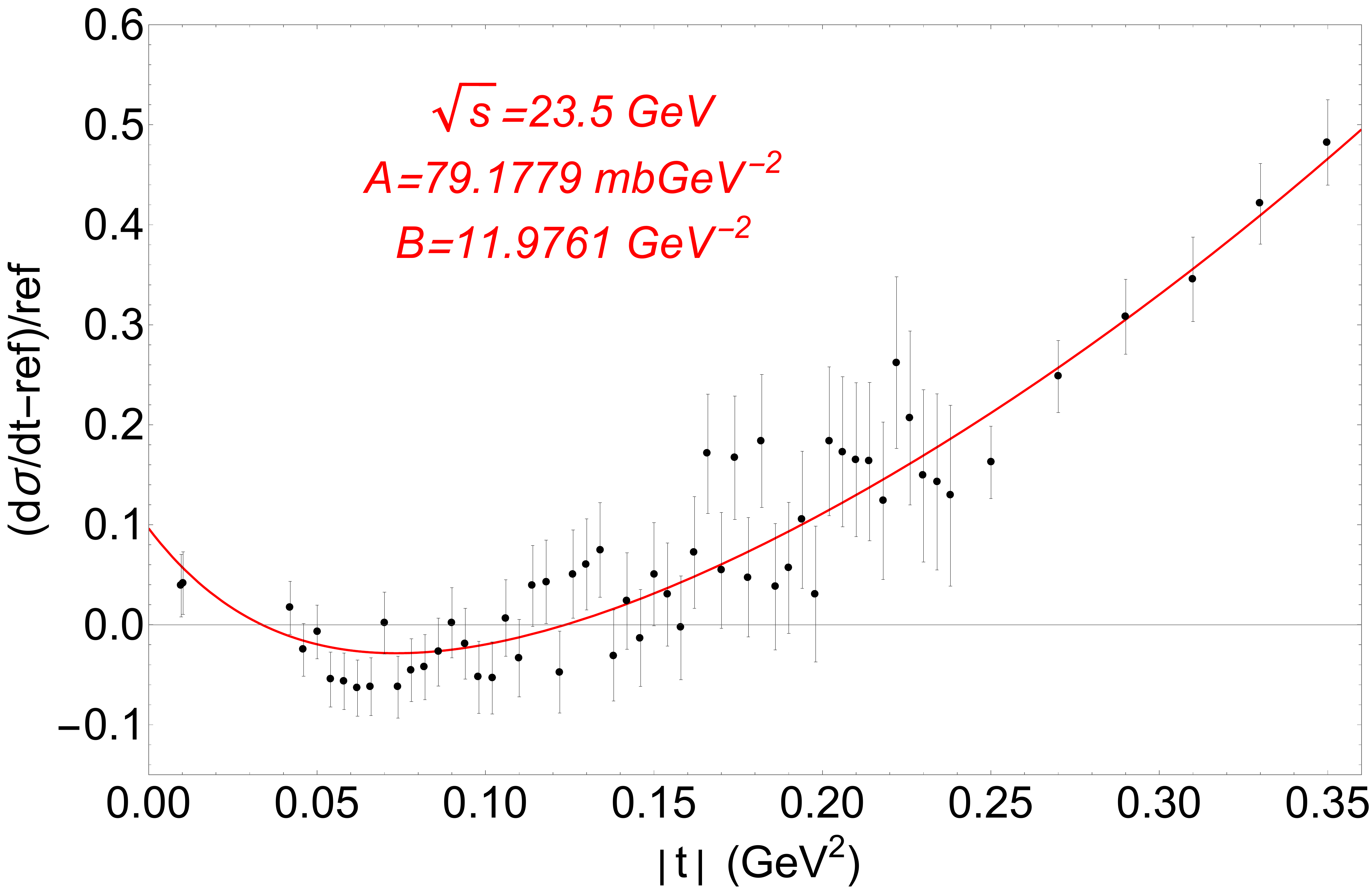}%
	}\hfill
	\subfloat[30.7 GeV\label{sfig:testa}]{%
		\includegraphics[scale=0.19]{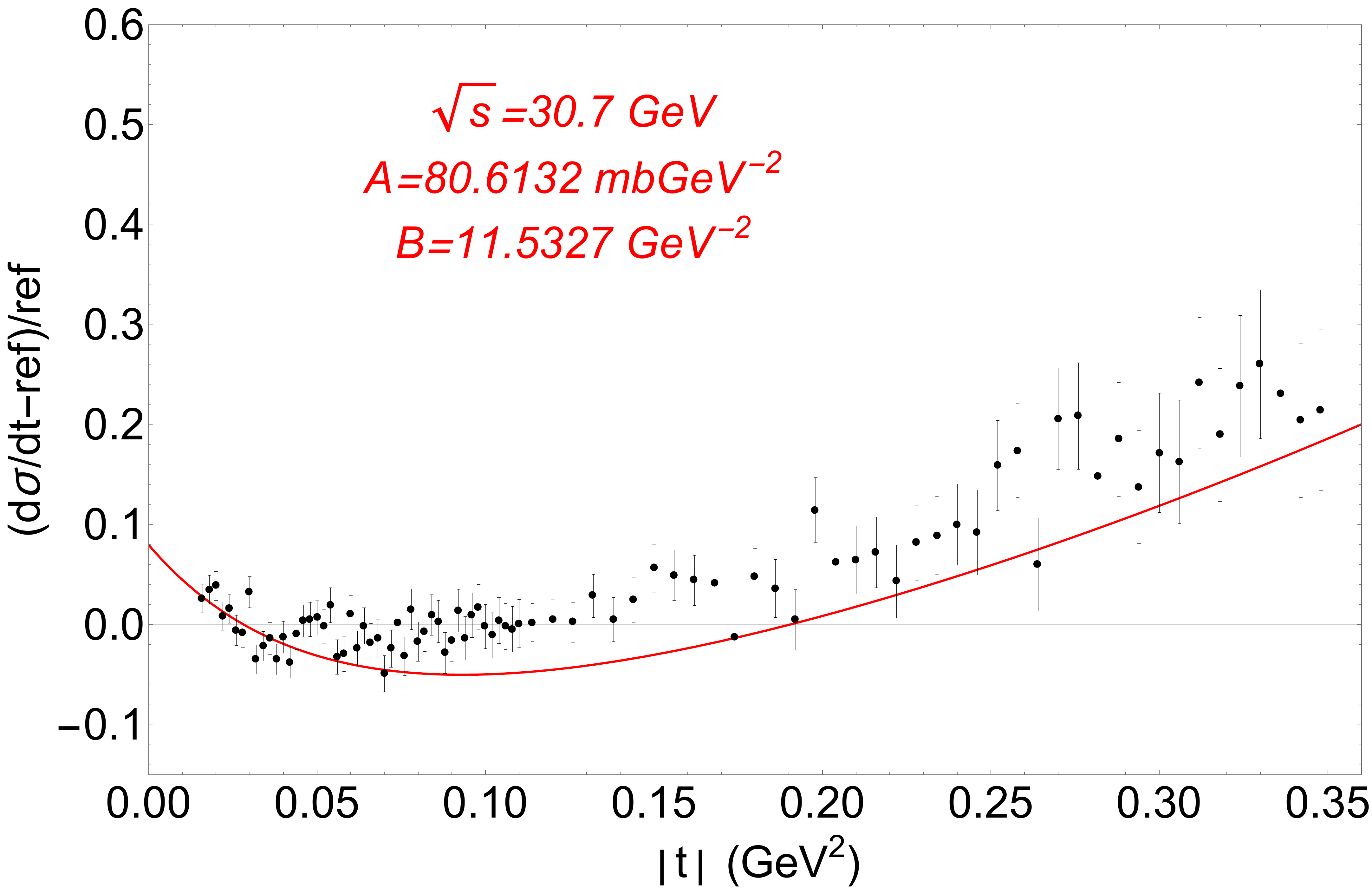}%
	}\hfill
	\subfloat[44.7 GeV\label{sfig:testa}]{%
		\includegraphics[scale=0.19]{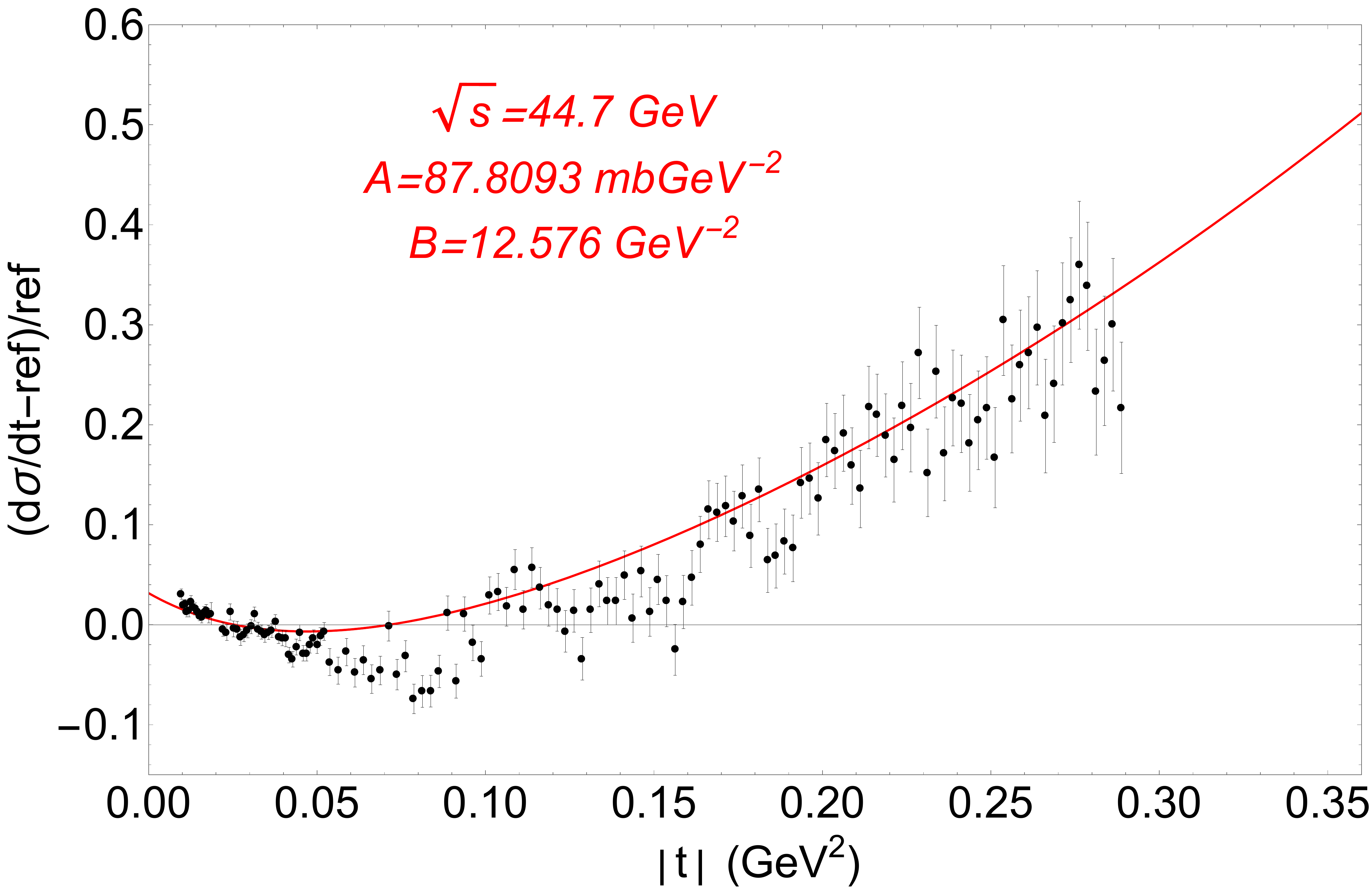}%
	}\hfill
	\subfloat[52.8 GeV\label{sfig:testab}]{%
		\includegraphics[scale=0.19]{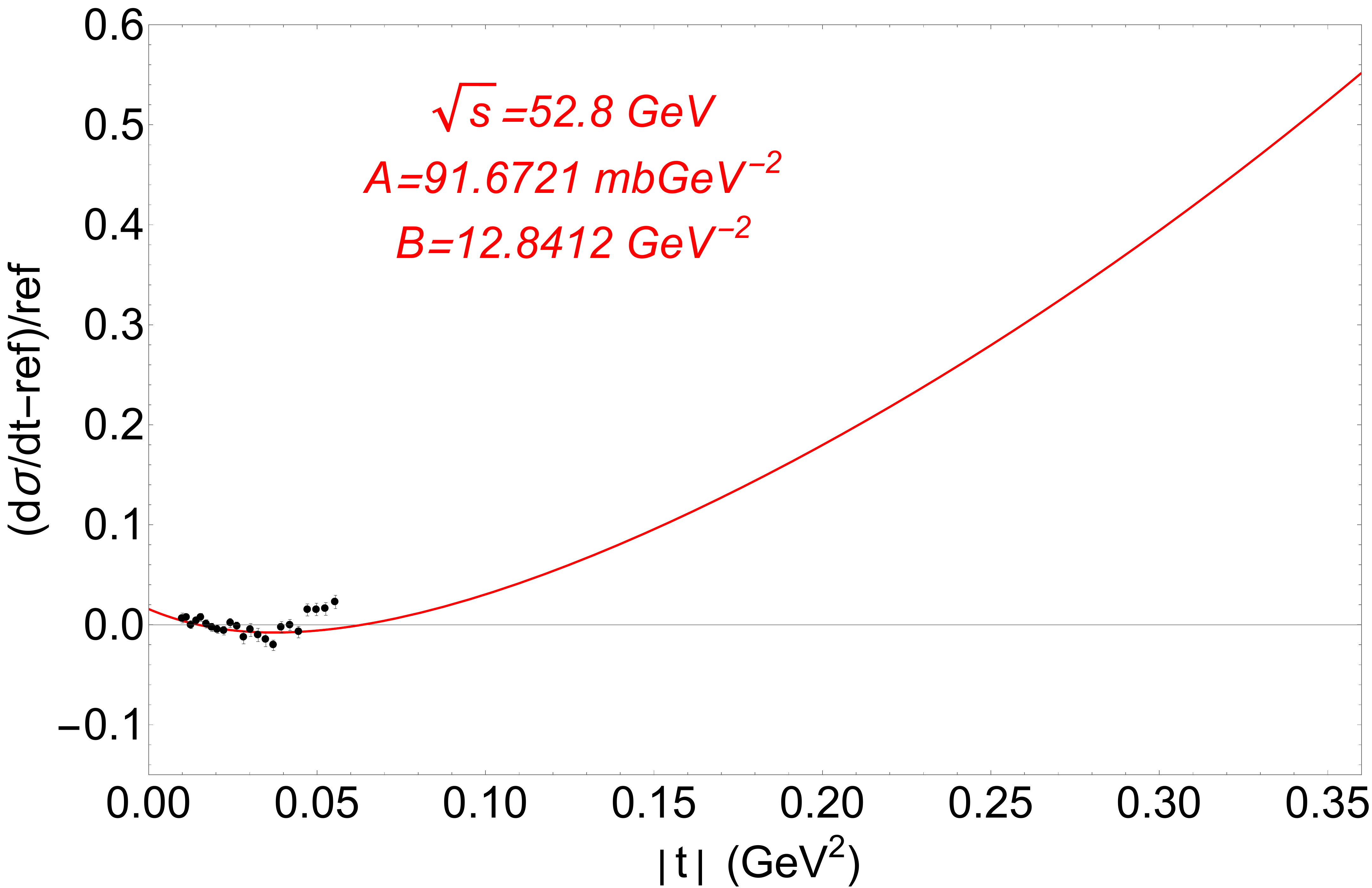}%
	}\hfill
	\subfloat[62.5 GeV\label{sfig:testab}]{%
		\includegraphics[scale=0.19]{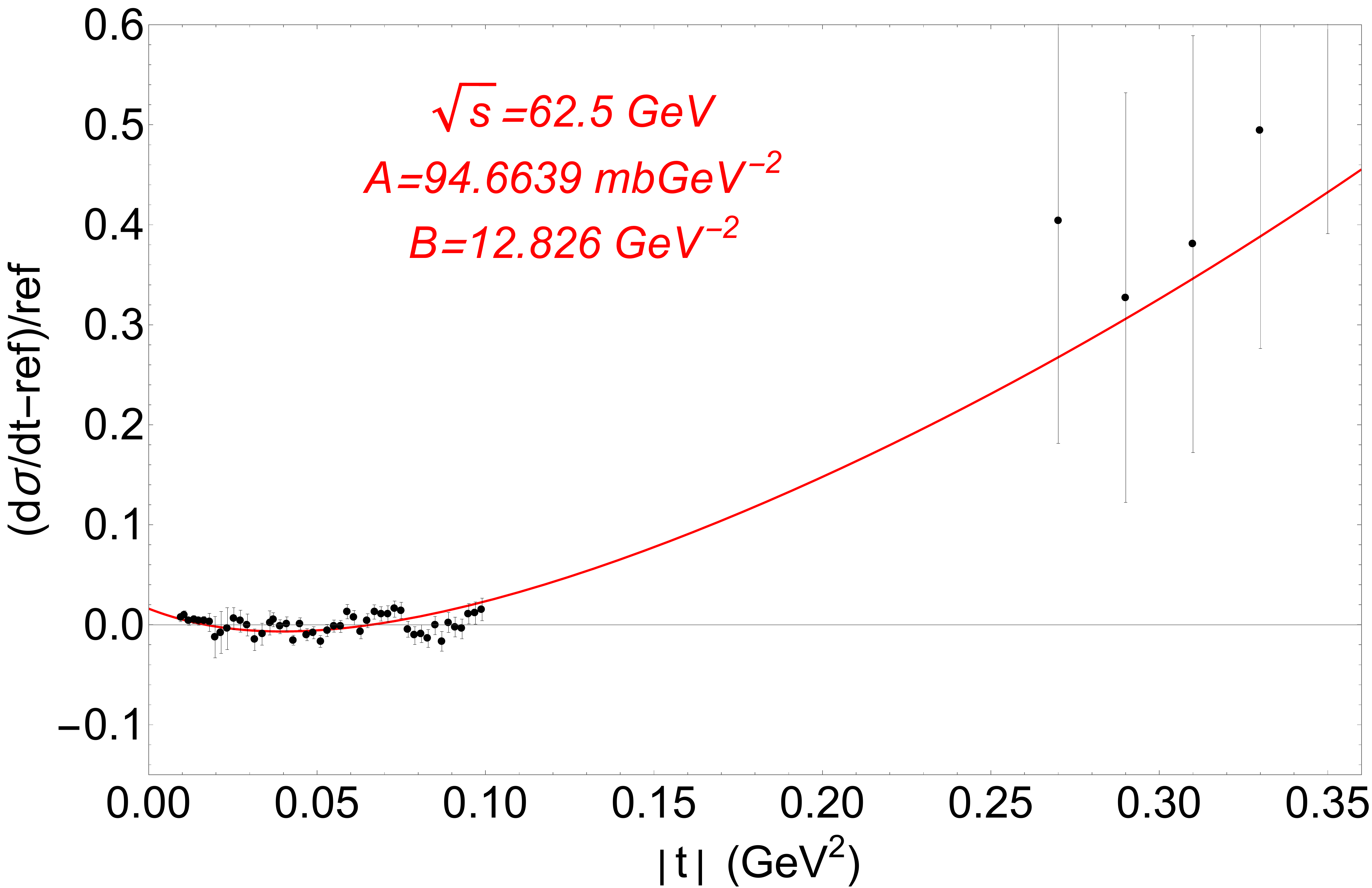}%
	}
	
	\caption{$R$ ratios calculated for the ISR \cite{ISR} energies with $4$ fitted parameters.}
	\label{Fig:ISR_normb}
\end{figure}

The local slope at the ISR, calculated as
\begin{equation}
B(s,t)=\frac{d}{dt}ln\frac{d\sigma}{dt}
\end{equation}
is shown in Fig. \ref{Fig:ISR_slopea} (in case of $10$ fitted parameters) and Fig. \ref{Fig:ISR_slopeb} (with $4$ fitted parameters) .
Anticipating the comparison with the LHC data in the next two sections, here we present the ISR data also in the normalized form, used by TOTEM as 
\begin{equation} \label{Eq:norm}
R=\frac{\frac{d\sigma}{dt}-ref}{ref},
\end{equation}
where $ref=Ae^{Bt}$. The result is shown in Fig. \ref{Fig:ISR_norma} (for $10$ fitted parameters) and Fig. \ref{Fig:ISR_normb} (with $4$ fitted parameters)

\section{The "break" at the LHC} \label{Sec:LHC}
\begin{figure}[H]
	\centering
	\includegraphics[scale=0.19]{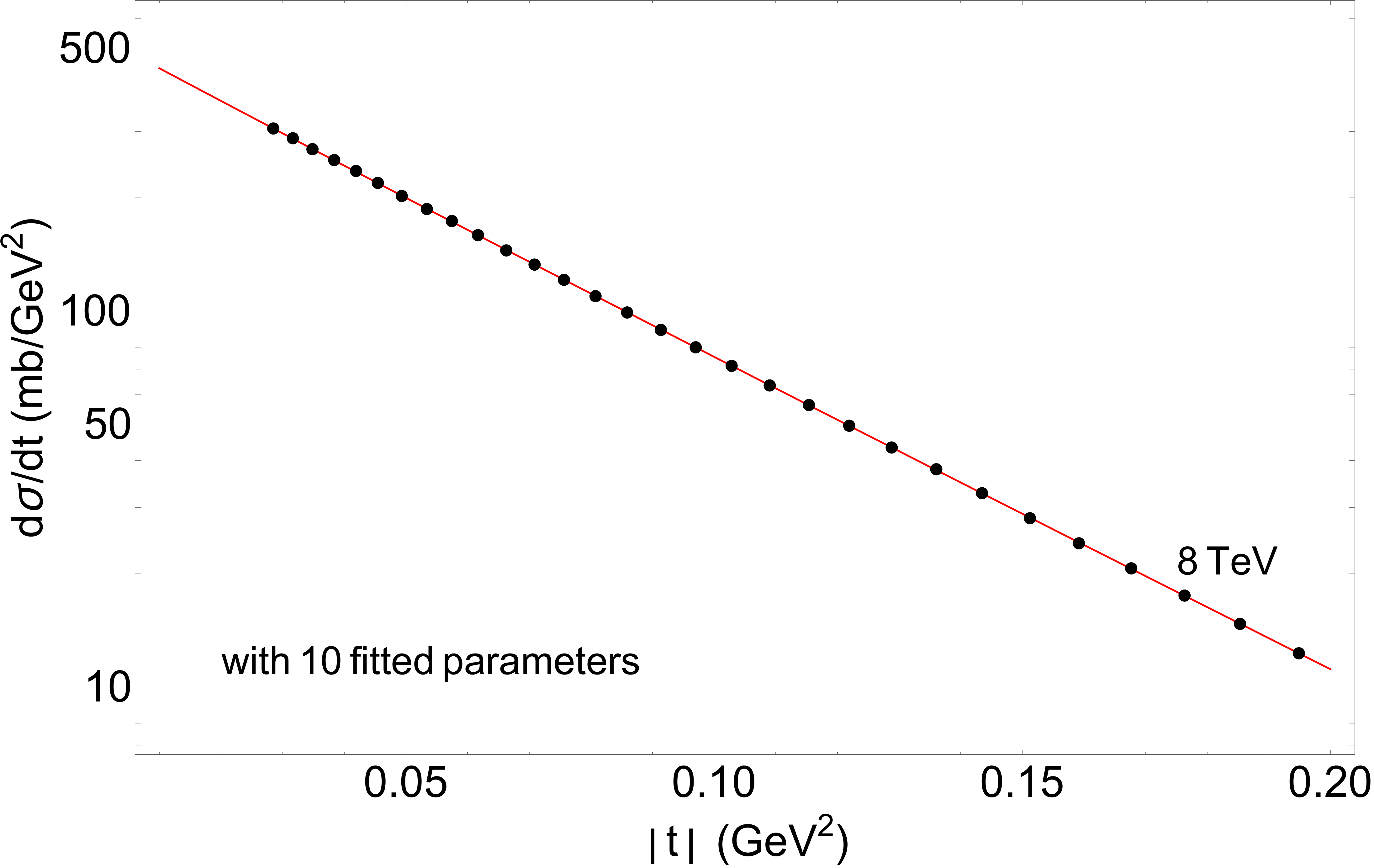}
	\includegraphics[scale=0.19]{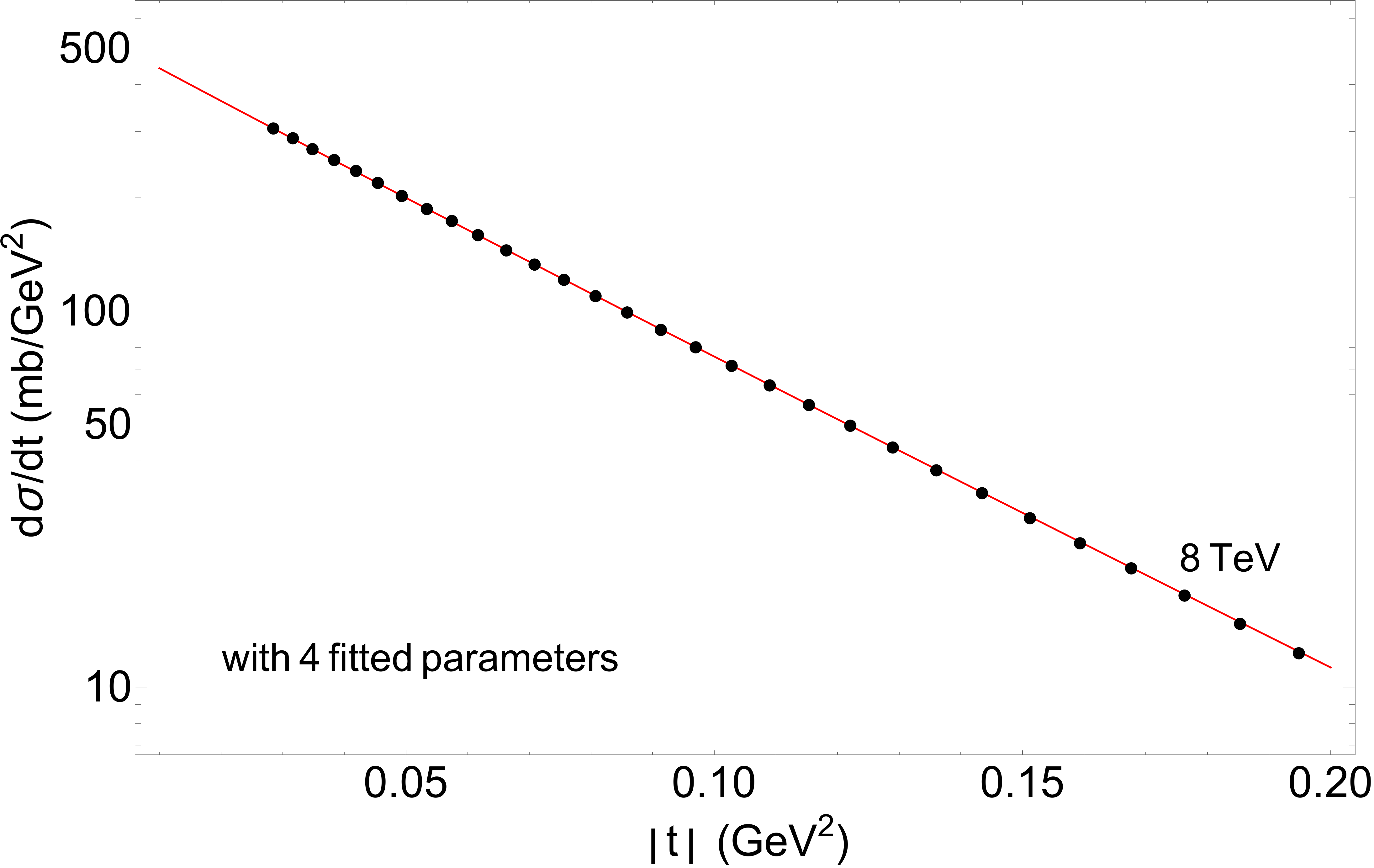}
	\caption{Result of the fit for TOTEM $8$ TeV data \cite{TOTEM8}.}
	\label{Fig:8TeV_fit}
\end{figure}

Now we proceed in the same way with the $8$ TeV TOTEM data. The results of the fits of the model Eqs. (\ref{Eq:ampl}),  (\ref{Eq:Pf}), (\ref{Eq:trajectory}) to the $8$ TeV data \cite{TOTEM8} are shown in Fig. \ref{Fig:8TeV_fit}.
The values of the fitted parameters are presented in Table II.
\begin{table}[H]
	\centering
	\subfloat[With 10 fitted parameters \label{sfig:testa}]{%
		\begin{tabular}{c c c c}\hline
			$\alpha_{0P}$ & 1.08893   &	$\alpha_{0f}$  & 0.614576\\
			$\alpha'_P$   & 0.463837  &$\alpha'_f$     & 0.953665\\
			$\alpha_{1P}$ & 0.0329217 &	$\alpha_{1f}$  &-0.0939313\\
			$a_P$         &1.18139   &$a_f$           &11.2719\\
			$b_P$         &1.16537   &$b_f$           &5.54316\\
			$s_{0P}$      &1 (fixed)  &$s_{0f}$        &1 (fixed)\\\hline
			&$\chi^2/DOF$ & 0.09845&\\
			&$DOF$ & 20& \\\hline
		\end{tabular}%
	}\qquad
	\subfloat[With 4 fitted parameters \label{sfig:testa}]{%
		\begin{tabular}{c c c c}\hline
			$\alpha_{0P}$ & 1.08 (fixed) &	$\alpha_{0f}$ & 0.5 (fixed)\\
			$\alpha'_P$   & 0.3 (fixed)  &$\alpha'_f$     & 1 (fixed)\\
			$\alpha_{1P}$ & 0.03 (fixed) &	$\alpha_{1f}$ &0.1 (fixed)\\
			$a_P$         &0.0000453477 &$a_f$           &117137\\
			$b_P$         & 10.7496      &$b_f$           &-20356.8\\
			$s_{0P}$      &1 (fixed)     &$s_{0f}$        &1 (fixed)\\\hline
			&$\chi^2/DOF$ & 0.09777&\\
			&$DOF$ & 26& \\\hline
		\end{tabular}%
	}
	
	\caption{ Values of the fitted parameters for the TOTEM $8$ TeV data \cite{TOTEM8}.}
	\label{}
\end{table}

\begin{figure}[H] 
	\centering
	\subfloat[Local slope\label{sfig:testa}]{%
		\includegraphics[scale=0.19]{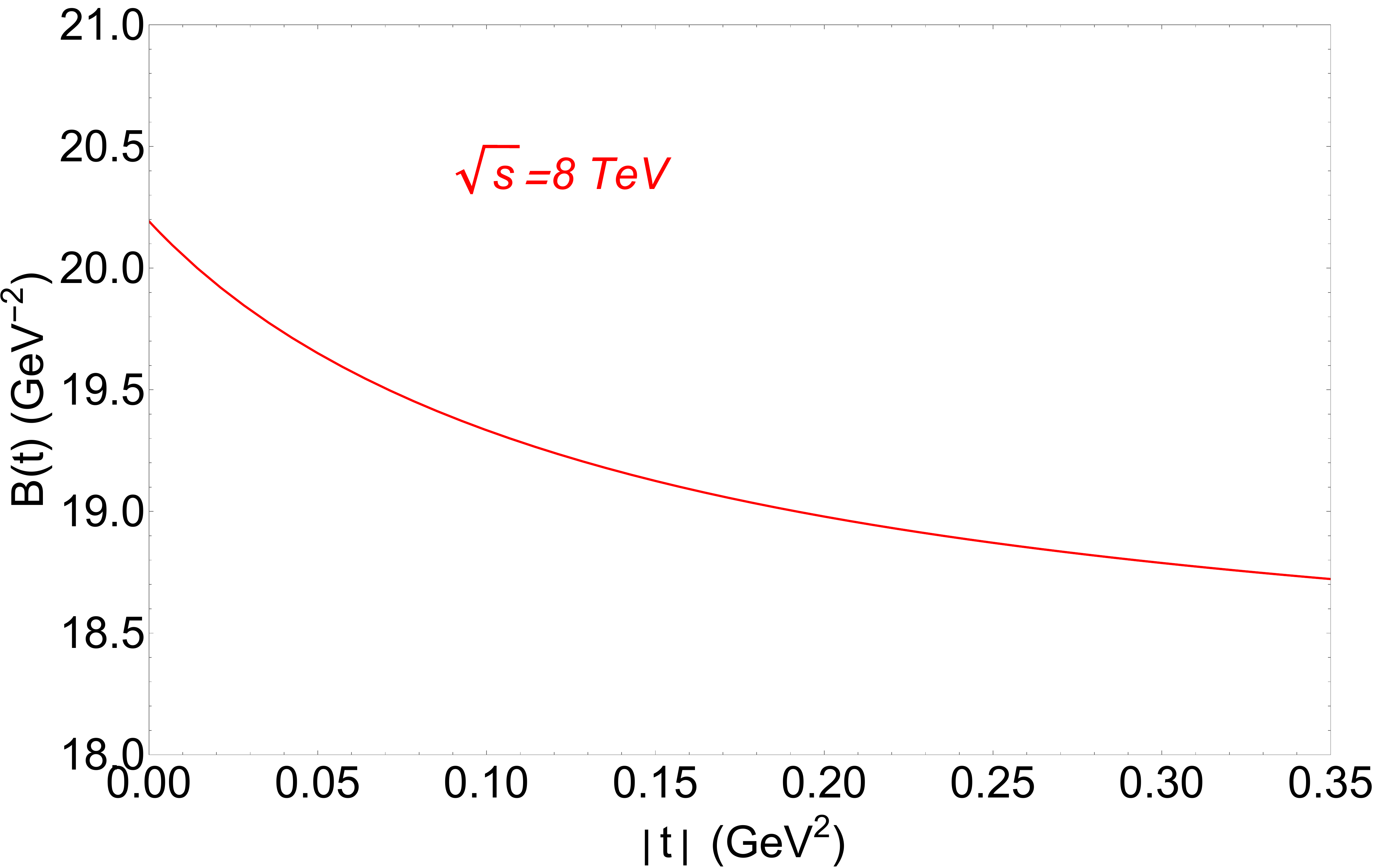}%
	}\quad
	\subfloat[R ratio\label{sfig:testa}]{%
		\includegraphics[scale=0.19]{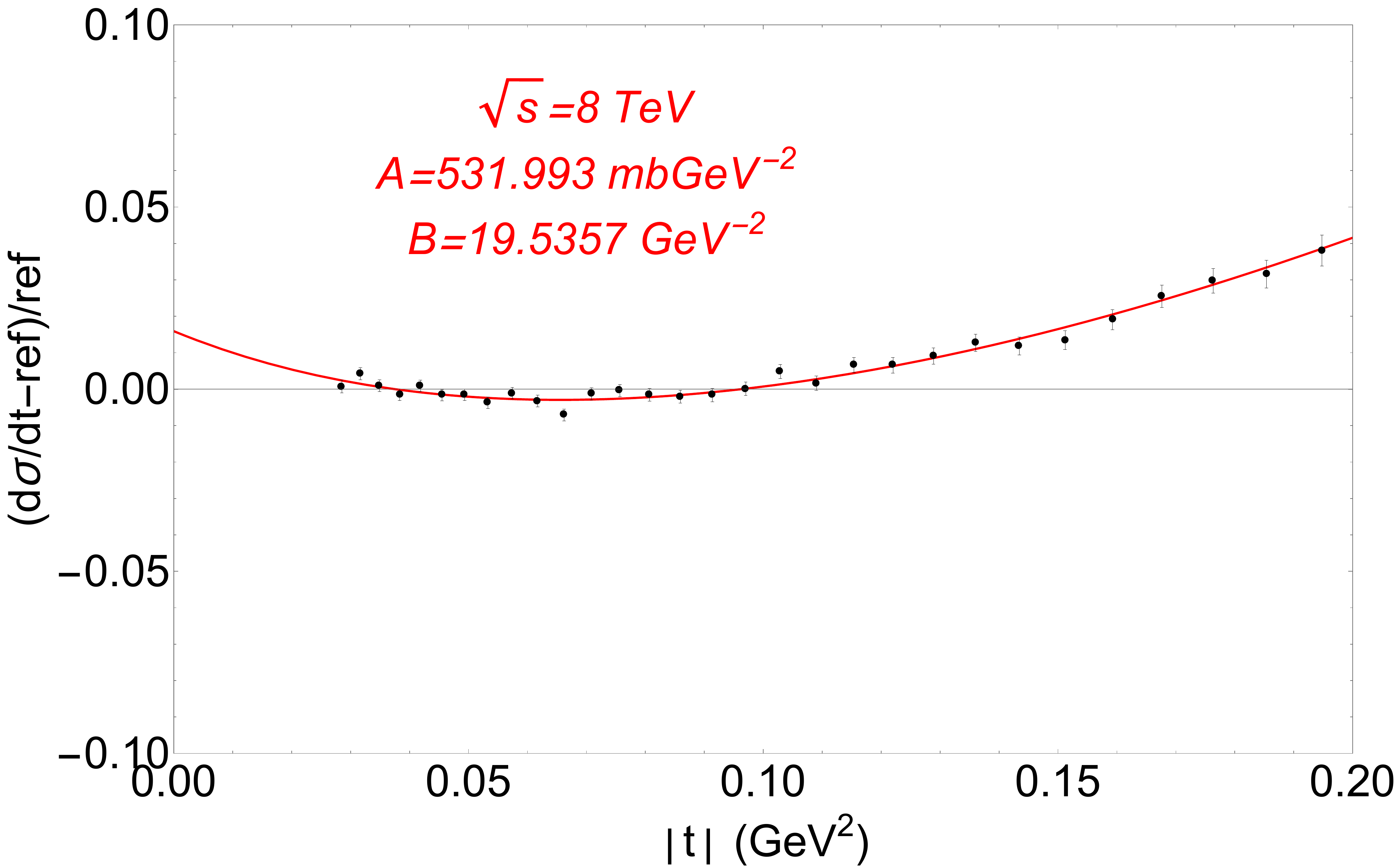}%
	}
	
	\caption{(a) Local slope and (b) the ratio $R$ calculated for the TOTEM $8$ TeV data \cite{TOTEM8} with $10$ fitted parameters.}
	\label{LHC_slopeRa}
\end{figure}

\begin{figure}[H] 
	\centering
	\subfloat[Local slope\label{sfig:testa}]{%
		\includegraphics[scale=0.19]{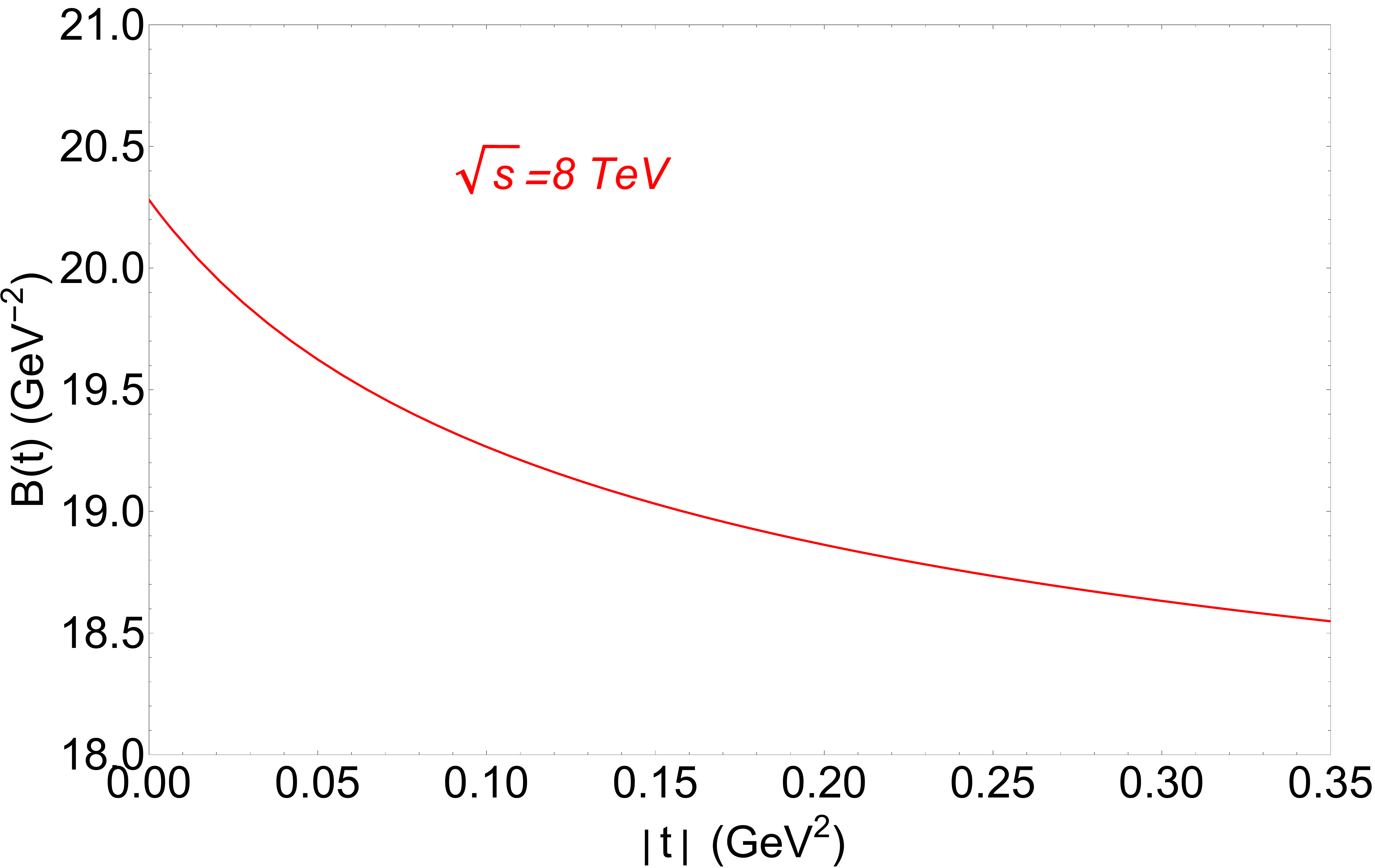}%
	}\quad
	\subfloat[R ratio\label{sfig:testa}]{%
		\includegraphics[scale=0.19]{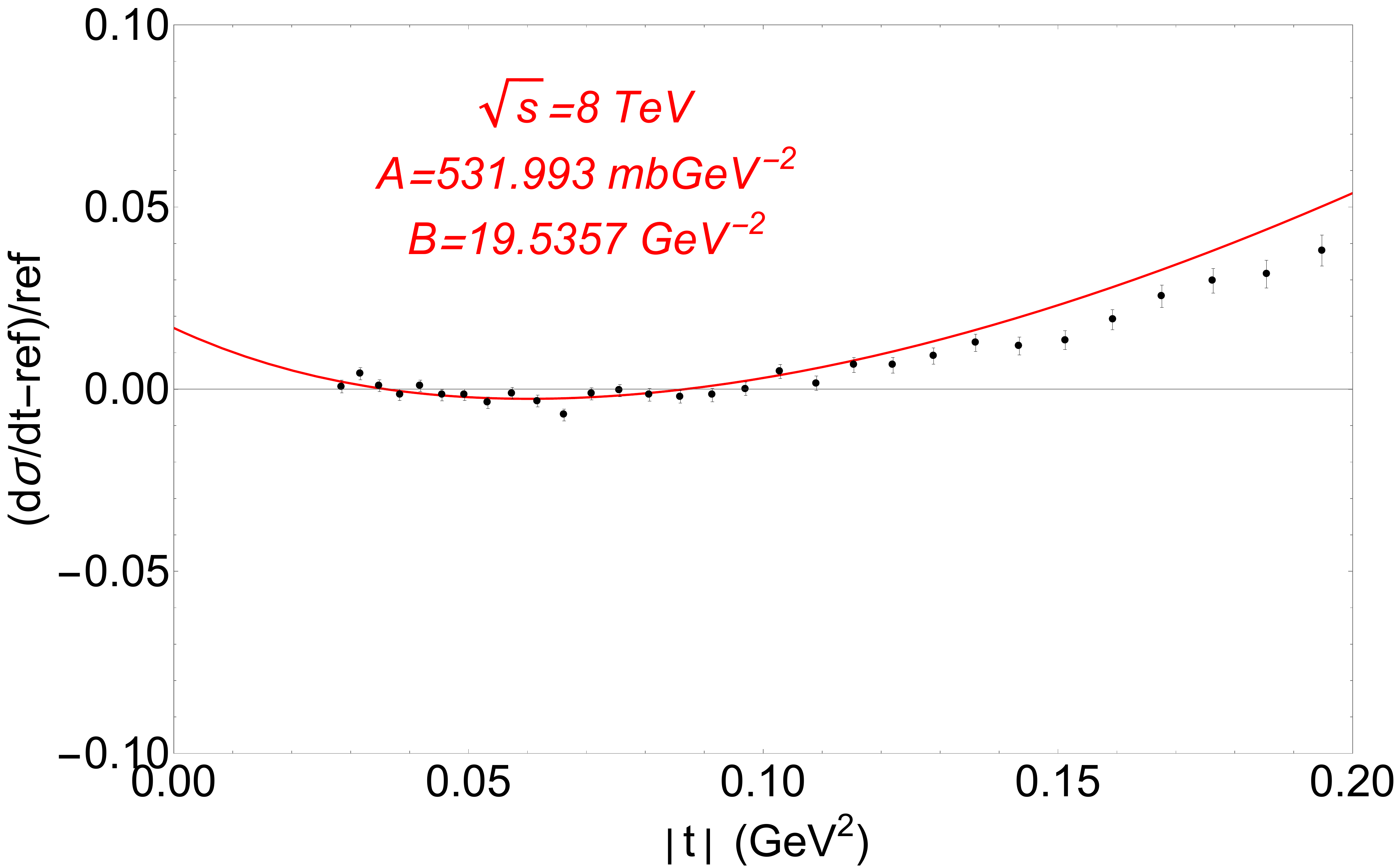}%
	}
	
	\caption{(a) Local slope and (b) the ratio $R$ calculated for the TOTEM $8$ TeV data \cite{TOTEM8} with $4$ fitted parameters.}
	\label{LHC_slopeRb}
\end{figure}

From these fits we calculate the relevant local slopes $B(s,t)$ and $R$ ratios for $8$ TeV. The results are shown in Fig. \ref{LHC_slopeRa} (with $10$ fitted parameters) and Fig. \ref{LHC_slopeRb} (with $4$ fitted parameters).


\begin{figure}[H] 
	\centering
	\subfloat[Differential cross section\label{sfig:testa}]{%
		\includegraphics[scale=0.19]{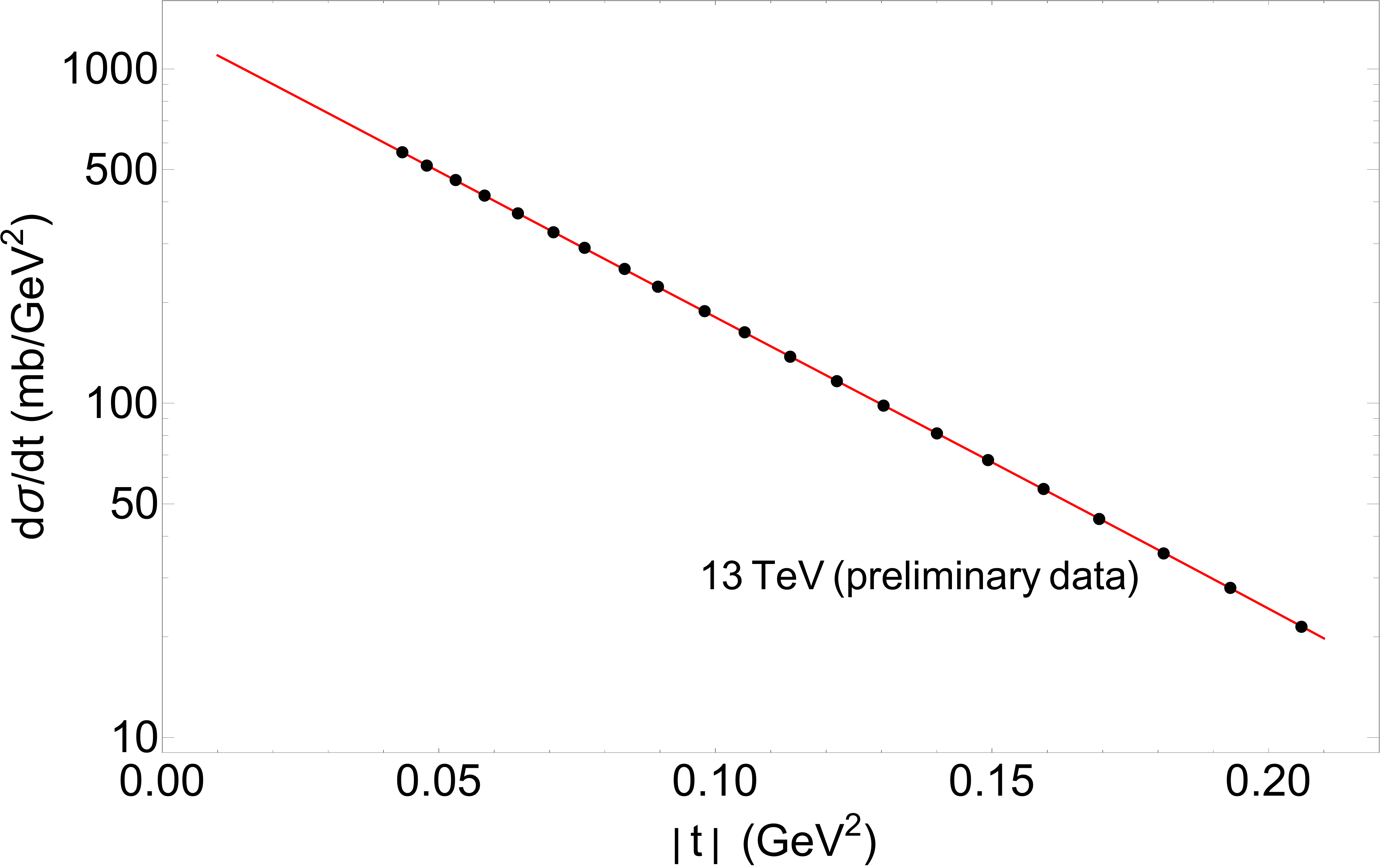}%
	}\quad
	\subfloat[R ratio \label{sfig:testa}]{%
		\includegraphics[scale=0.19]{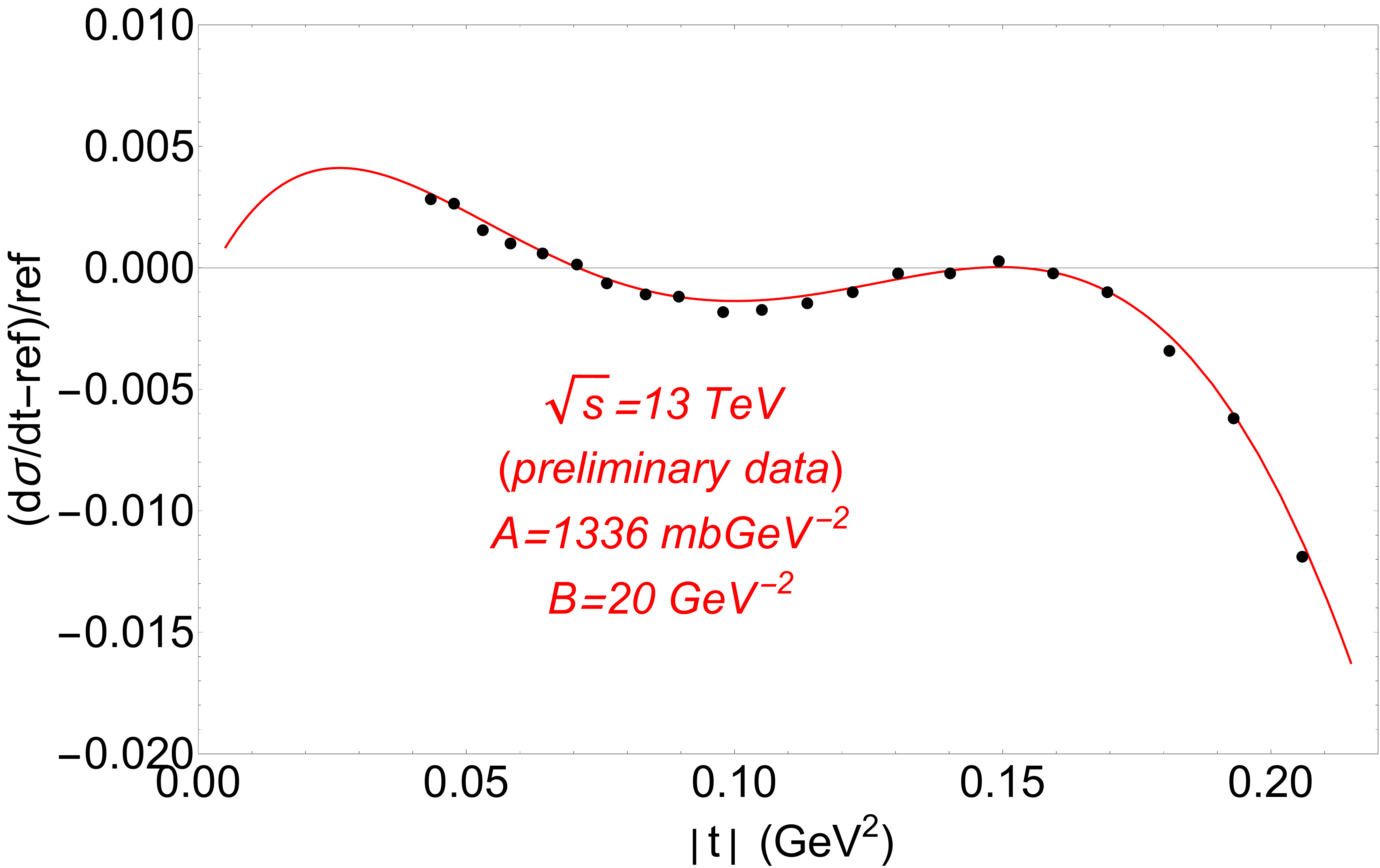}%
	}
	
	\caption{Predictions for (a) differential cross section and (b) R ratio at $13$ TeV.}
	\label{Fig:Predict}
\end{figure}

Predictions for the differential cross section and $R$ ratio at $13$ TeV are shown in Fig. \ref{Fig:Predict}.


\section{Mapping the "low-energy" break to that at the LHC} \label{Extrapolate}
\begin{figure}[H] 
	\centering
	\includegraphics[scale=0.19]{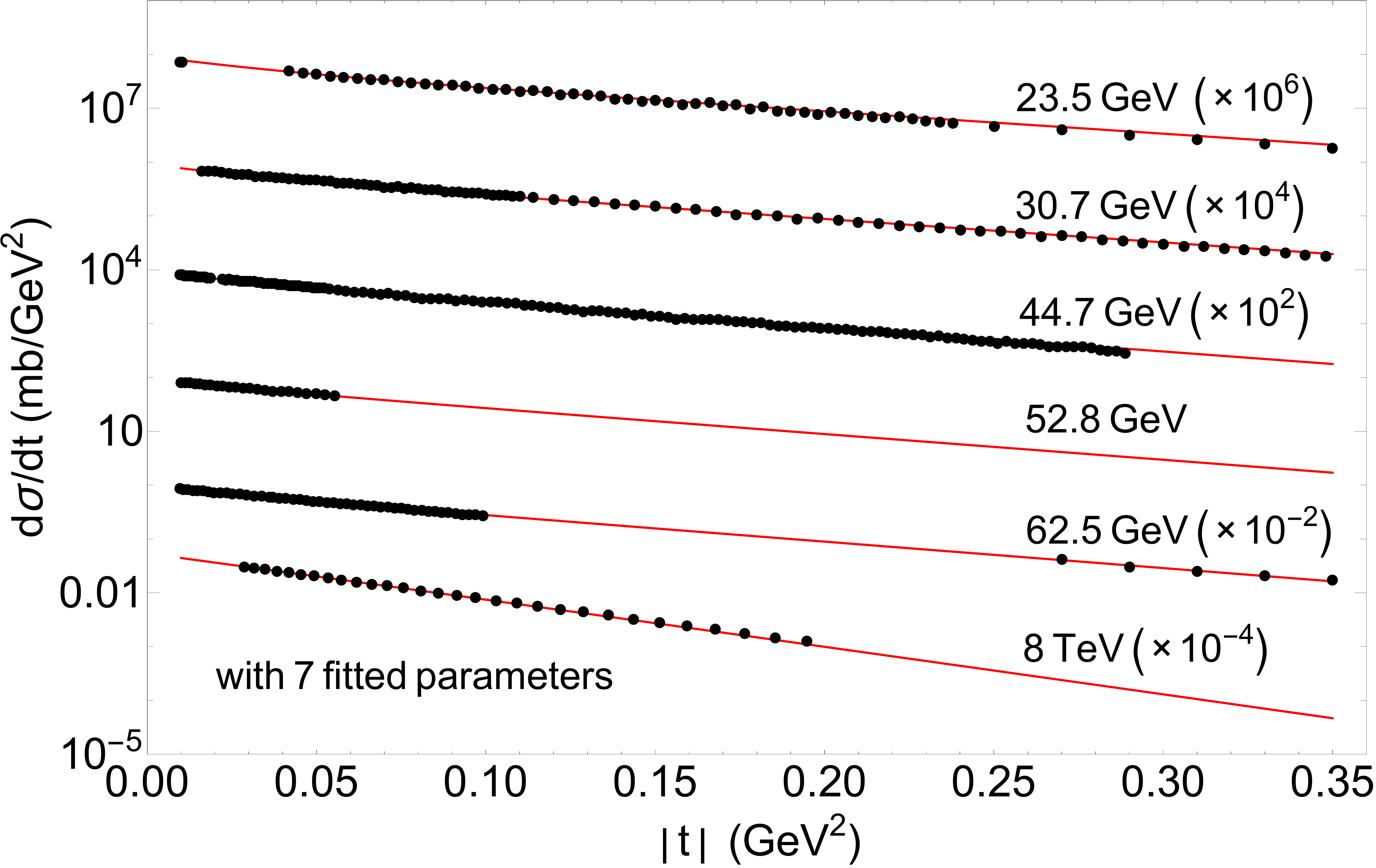}
	\includegraphics[scale=0.19]{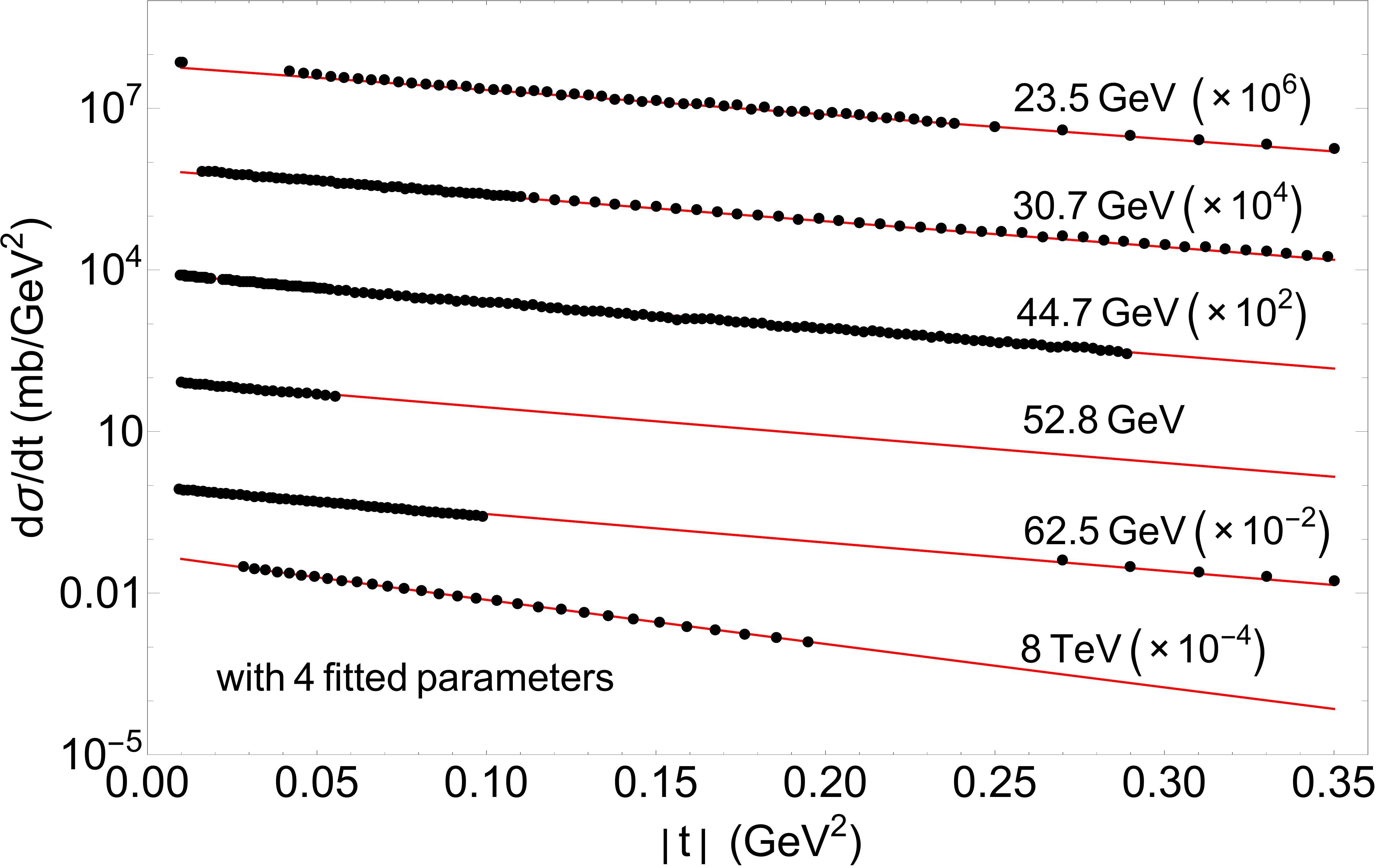}
	\caption{Result of our extrapolation using the ISR and TOTEM $8$ TeV data \cite{TOTEM8}.}
	\label{Fig:Map}
\end{figure}

By using the Regge pole model, Eqs. (\ref{Eq:ampl}),  (\ref{Eq:Pf}) and (\ref{Eq:trajectory}), now we  map the "break" fitted at the ISR onto the TOTEM data. The result is shown in Fig. \ref{Fig:Map}.

The values of the fitted parameters are presented in Table III.
\begin{table}[H]
	\centering
	\subfloat[With 7 fitted parameters \label{sfig:testa}]{%
		\begin{tabular}{c c c c}\hline
			$\alpha_{0P}$ & 1.0971   &	$\alpha_{0f}$  & 0.5 (fixed)\\
			$\alpha'_P$   & 0.487269  &$\alpha'_f$     & 1 (fixed)\\
			$\alpha_{1P}$ &-0.0242427 &	$\alpha_{1f}$  &0.1 (fixed)\\
			$a_P$         &0.0538938  &$a_f$           &0.0735019\\
			$b_P$         &3.84255   &$b_f$           &13.6625\\
			$s_{0P}$      &1 (fixed)  &$s_{0f}$        &1 (fixed)\\\hline
			&$\chi^2/DOF$ &1.0033&\\
			&$DOF$ & 386& \\\hline
		\end{tabular}%
	}\qquad
	\subfloat[With 4 fitted parameters \label{sfig:testa}]{%
		\begin{tabular}{c c c c}\hline
			$\alpha_{0P}$ & 1.08 (fixed) & $\alpha_{0f}$   & 0.5 (fixed)\\
			$\alpha'_P$   & 0.3 (fixed)  & $\alpha'_f$     & 1 (fixed)\\
			$\alpha_{1P}$ & 0.03 (fixed) & $\alpha_{1f}$   & 0.1 (fixed)\\
			$a_P$         & 0.0000594678 & $a_f$           & -28.6249\\
			$b_P$         & 10.4853      & $b_f$           & 0.882867\\
			$s_{0P}$      & 1 (fixed)    & $s_{0f}$        & 1 (fixed)\\\hline
			&$\chi^2/DOF$ & 8.8174       & \\
			&$DOF$ & 389& \\\hline
		\end{tabular}%
	}
	
	\caption{ Values of fitted parameters in  our extrapolation from ISR to the LHC.}
	\label{sfig:testa}
\end{table}

\section{Conclusions} \label{Sec:Conclude}
The successful fit to the proton-proton total cross section with a simple model, Figs. \ref{Fig:Total1} and \ref{Fig:Total2} shows the efficiency of Regge poles in reproducing energy dependence. Much more complicated is the parametrization of the $t$-dependence, containing irregularities, in particular the "break" under discussion. Still, most of the fits presented in this paper are coherent and differ only by small details. To make the picture complete, we have quoted several options; one can see that the results are not too sensitive to the choice of the number of free parameters. For a better comparison between the "break" as seen at ISR and that at the LHC, we have refitted the ISR data, normalizing to an exponential "test function" (Figs. \ref{Fig:ISR_norma} and \ref{Fig:ISR_normb}) as done at the LHC \cite{TOTEM8, TOTEM13}. The resulting predictions, Fig. \ref{Fig:Predict} for the LHC energy $13$ TeV may be of interest for experimentalists.  

We have shown that the deviation from a linear exponential of the $pp$ diffraction cone as seen at the ISR, $20.3\leq\sqrt s\leq 62.5$ GeV and at the LHC, $\sqrt s=8$ and $13$ TeV are of  similar nature: they appear nearly at the same value of $t\approx -0.1$ GeV$^2$, have the same concave shape of comparable "size", $\Delta B(t)\approx 2\div 4$ GeV$^{-2}$ and may be fitted by similar $t$-dependent function. Mapping this $t$-dependence through the tremendous energy span from the ISR to the LHC (almost 3 orders of magnitude) is a highly non-trivial task. We have done it within the simplest Regge pole model, with two trajectories: a leading one, the Pomeron and a sub-leading effective Reggeon. More advanced and refined Regge-type models may improve the fit and clarify details.   
    
The threshold singularity in question should be present also in the $f$ trajectory, however it has secondary effect with respect to the Pomeron. 

Note also that the low-$|t|$ structure of the diffraction cone was fitted also \cite{Lia} by a relevant form factor (Regge residue).

The results presented in this paper leave open and raise also several questions, namely:

1) theoretical calculations of the relative weight of the loop contribution, second term in Fig. \ref{Fig:Diagram} relative to the first one ("Born term") are needed; 

2) why is the "break" observed only in elastic $pp$ scattering, not in $p\bar p,$ for example at the Tevatron? Once the Pomeron is universal, the effect should be present also in $p\bar p$. Non-observation of any convex or concave curvature in the diffraction cone at the Tevatron may be attributed to poor statistics of the relevant data (lacking Roman pots), preventing the observation of such a tiny effect.

To conclude, we expect more precise data in the low-$|t|$ region on elastic scattering and diffraction dissociation as well as further fits with improved phenomenological parametrizations. Theoretical calculations of the diagram (Fig. \ref{Fig:Diagram}) may shed more light on the nature of the phenomenon. Needless to say, further attempts in this direction will be based on improved models for the scattering amplitude, with more details on individual Regge trajectories, including the Odderon. 

\subsection*{Acknowledgements}
L. J. was supported by the Program "Matter under Extreme Conditions" of the Nat. Ac. Sc. of Ukraine.


\end{document}